\documentclass[10pt,aps,prb,superscriptaddress,twocolumn,amsfonts,amsmath,a4paper,showpacs]{revtex4}
\usepackage{graphicx}

\begin{document}

\title{Relaxation Scenarios in a Mixture of Large and Small Spheres:\\ 
Dependence on the Size Disparity}

%
%
\author{Angel J. Moreno}
\email[Corresponding author: ]{wabmosea@sq.ehu.es}
\affiliation{Donostia International Physics Center, Paseo Manuel de Lardizabal 4,
20018 San Sebasti\'{a}n, Spain.}
\author{Juan Colmenero}
\affiliation{Donostia International Physics Center, Paseo Manuel de Lardizabal 4,
20018 San Sebasti\'{a}n, Spain.}
\affiliation{Departamento de F\'{\i}sica de Materiales, Universidad del Pa\'{\i}s Vasco (UPV/EHU),
Apartado 1072, 20080 San Sebasti\'{a}n, Spain.}
\affiliation{Unidad de F\'{\i}sica de Materiales, Centro Mixto CSIC-UPV, 
Apartado 1072, 20080 San Sebasti\'{a}n, Spain.}

\begin{abstract}

We present a computational investigation on the slow dynamics of a mixture
of large and small soft spheres. By varying the size disparity at a moderate fixed composition
different relaxation scenarios are observed for the small particles.
For small disparity density-density correlators exhibit moderate stretching.
Only small quantitative differences are observed between dynamic features for
large and small particles. On the contrary, large disparity induces a clear time scale separation
between the large and the small particles. Density-density correlators
for the small particles become extremely stretched, and display
logarithmic relaxation by properly tuning the temperature or the wavevector.
Self-correlators decay much faster than density-density correlators.
For very large size disparity, a complete separation between self- and collective dynamics
is observed for the small particles. Self-correlators decay to zero at temperatures where density-density
correlations are frozen. The dynamic picture obtained by varying the size disparity
resembles features associated to Mode Coupling transition lines of the types B and A
at, respectively, small and very large size disparity. Both lines might merge, at some
intermediate disparity, at a higher-order point, to which logarithmic relaxation
would be associated. This picture resembles predictions of a recent
Mode Coupling Theory for fluids confined in matrixes with interconnected voids
[V. Krakoviack, Phys. Rev. Lett. {\bf 94}, 065703 (2005)].

\end{abstract}
\date{\today}
\maketitle
\newpage
\begin{center}
\bf{I. INTRODUCTION}
\end{center}

Relaxation dynamics in glass-forming systems can be strongly sped up or slowed down by the addition
of a component of very different mobility. An example is provided by mixtures
of large and small particles of very different size. Several investigations
have evidenced the possibility of melting a glass of large particles by adding a proper amount
of small particles \cite{williams,eckert,starpol}.
Less attention has been paid to the dynamic features exhibited by the small particles.
Very recently, we have presented a molecular dynamics investigation on the relaxation
dynamics of a mixture of large and small soft spheres \cite{mixturepaper}.
We define the ``size disparity'', $\delta$, as the ratio of the diameters
of the large and the small particles.
For a sufficiently large disparity ($\delta = 2.5$ in Ref. \cite{mixturepaper}), 
at moderate and low concentrations of the small particles,
the latters exhibit unusual relaxation features.
Differently from the usual two-step increase and decay observed for, respectively,
mean squared displacements and dynamic correlators \cite{koband,kobrev,mctrev1,mctrev2}, 
these quantities do not exhibit a defined plateau at intermediate times between
the microscopic and diffusive regimes \cite{mixturepaper}.
This result suggests a softer character for the collective caging mechanism --- i.e., the temporary
trapping of each particle by its neighbors. 
Dynamic correlators show a highly stretched decay, and for selected values of the control
parameters the decay is logarithmic in time. By varying wavevectors or control parameters
the decay shows a striking concave-to-convex crossover \cite{mixturepaper}.

These anomalous relaxation features resemble predictions 
of the Mode Coupling Theory (MCT) \cite{mctrev1,mctrev2,bengtzelius,das,reichman}
for state points close to higher-order MCT transitions, initially predicted by schematic models \cite{schematic}, 
and later derived for one-component systems of particles interacting through a repulsive potential
complemented by a short-ranged attraction \cite{fabbian,bergenholtz,dawson,sperl,jpcmA4}.
These models are used as effective pictures for the {\it large} particles. The short-ranged attraction
arises as an effective interaction between the large particles, originating from
the depletion mechanism induced by the presence of the small particles \cite{asakura,likos}. 
In a certain range of density, temperature and mixture composition,
competition occurs between two different mechanisms for dynamic arrest of the large particles:
steric repulsion characteristic of colloidal systems, and formation of reversible
bonds, induced by the effective short-ranged attraction. 
The higher-order MCT scenario arises as a consequence of these two competing mechanisms
of very different localization lengths \cite{sperl,dawson}.
When heating up or cooling down the system, dynamic arrest is exclusively driven
by, respectively, steric repulsion and reversible bond formation, and relaxation features
of standard liquid-glass transitions are recovered \cite{prlsimA4,zaccarelli1}.
Many of the predictions associated with the higher-order MCT scenario for short-ranged
attractive colloids has been succesfully tested by simulations 
\cite{prlsimA4,zaccarelli1,zaccarelli2,foffiA4,puertas,sciotart}
and experiments \cite{mallamace,eckert,sciotart,pham,chen,pontoni,grandjean}.

If the size disparity is sufficiently large to induce a large time scale separation
between large and small particles, mean squared displacements and density-density 
correlators for the small particles in the mixture of Ref. \cite{mixturepaper} 
display striking similarities with qualitative features
associated to higher-order MCT transitions. Similar results have been reported
for the fast component in simulations of polymer blends with components
of very different mobilities \cite{blendpaper,genix}.
These analogies with relaxation features 
in short-ranged attractive colloids
suggest that the higher-order MCT scenario might be a general feature
of systems showing a competition between different mechanisms for dynamic arrest.
For the case of the small particles in the mentioned polymeric and non-polymeric mixtures,
we have suggested a competiton between bulk-like caging and confinement \cite{mixturepaper,blendpaper}.
Bulk-like caging is induced by neighboring small particles and confinement is induced by
the slow matrix formed by the large particles.

In this article we present new support for this interpretation by evidencing dynamic analogies
with recent MCT theoretical calculations by Krakoviack \cite{krakoviackprl, krakoviackjpcm}
in a mixture of {\it mobile and fixed} particles, which explicitely shows a higher-order MCT transition.
We have carried out new simulations, at a fixed mixture composition, 
for disparities $1 \lesssim \delta \le 8$, extending results reported in Ref. \cite{mixturepaper}
for $\delta = 2.5$. For $\delta \rightarrow 1$ standard relaxation features are recovered.
For the largest investigated values of $\delta$ a new relaxation 
scenario arises for the small particles,
showing features characteristic of nearby MCT transitions of the so called type-A. 
Such transitions are defined by a zero value of the long-time limit (critical non-ergodicity parameter)
of density-density correlators, different from the finite value (observed at smaller $\delta$)
defining the usual type-B transitions.
This feature agrees with early MCT predictions by Bosse and co-workers \cite{bosse,kaneko}
and experimental work by Imhof and Dhont \cite{imhof}
for binary mixtures of hard spheres with very large size disparity.

Having mentioned these antecedents, simulations presented here constitute, to the best of our knowledge,
the first systematic investigation of slow relaxation 
(by analyzing mean squared displacements and dynamic correlators)
in binary mixtures covering the {\it whole} range between the limits of small and large disparity.
They also provide a connection with results in Refs. \cite{krakoviackprl, krakoviackjpcm},
which report a dynamic phase diagram displaying an A- and a B-line merging at a higher-order point.
We suggest that the anomalous relaxation features observed at moderate disparity 
$\delta \sim 2.5$ for the system here investigated
might originate from the existence of a nearby B-line (yielding finite values 
for the non-ergodicity parameters) ending at a nearby higher-order point, 
to which anomalous relaxation features would be associated. The B- and A-lines would 
extend from the higher-order point to, respectively, the limits of small ($\delta = 1$) 
and large size disparity. 

The article is organized as follows. In Section II we introduce the investigated model and
give computational details. In Section III we present simulation results for static 
structure factors, mean squared displacements and dynamic correlators. In Section IV we discuss
dynamic quantities in the framework of the MCT and propose a picture for the different relaxation
scenarios observed by varying the size disparity. Conclusions are given in Section V.

\begin{center}
\bf{II. MODEL AND SIMULATION DETAILS}
\end{center}

As in the work of Ref. \cite{mixturepaper}, we have simulated a mixture of large and small particles
of equal mass $m=1$, interacting through a soft-sphere potential plus a quadratic term:
\begin{equation}
V_{\alpha\beta} = 4\epsilon \left [\left (\frac{\sigma_{\alpha\beta}}{r}\right )^{12}
- C_0 + C_2\left (\frac{r}{\sigma_{\alpha\beta}}\right )^{2}\right ],
\label{eq:potsoft}
\end{equation}
where $\epsilon=1$, and $\alpha$, $\beta$ $\in$ \{A, B, C, D\}.
Two sets of large particles (labelled as A and B) and of small particles (C and D) are
introduced in order to avoid crystallization effects (see below).
The interaction is zero beyond a cutoff distance $c\sigma_{\alpha\beta}$, with $c = 1.15$.
The addition of the quadratic term to the soft-sphere interaction, with
the values $C_0 = 7c^{-12}$ and $C_2 = 6c^{-14}$, guarantees continuity of potential and forces at the
cutoff distance.
The diameters of the soft-sphere potential for the different types of interaction are:
$\sigma_{\rm DD} =1$, $\sigma_{\rm CC} = 1.1\sigma_{\rm DD}$,  
$\sigma_{\rm BB} = \delta\sigma_{\rm DD}$, $\sigma_{\rm AA} = 1.1\sigma_{\rm BB}$,
and $\sigma_{\alpha\beta}=(\sigma_{\alpha\alpha}+\sigma_{\beta\beta})/2$ for the case $\alpha \ne \beta$.
We have investigated the size disparities $\delta =$ 1.15, 1.6, 2.5, 5, and 8. 

The potential (\ref{eq:potsoft}) is purely repulsive. It does not show local minima within 
the interaction range $r < c\sigma_{\alpha\beta}$. Hence, slow dynamics in the present model
arises as a consequence of steric effects. MCT theoretical works are usually carried out 
on systems of hard objects, while simulations in similar systems with continuous interactions
are usually preferred for computational simplicity.  In the present system, 
the tail of the interaction potential is progressively probed by decreasing temperature, 
which qualitatively plays the role of increasing packing in a system of hard spheres. 

The mixture composition is defined as the fraction of small particles: 
$x_{\rm small} = (N_{\rm C} + N_{\rm D})/(N_{\rm A} + N_{\rm B} + N_{\rm C}+N_{\rm D} )$,
with $N_{\alpha}$ denoting the number of particles of the species $\alpha$.
We investigate a single composition $x_{\rm small} = 0.6$. 
The number of large and small particles is, respectively, 
$N_{\rm A}+N_{\rm B} = 1000$ and $N_{\rm C}+N_{\rm D} = 1500$.
We impose the constraints $N_{\rm A}=N_{\rm B}$ and $N_{\rm C}=N_{\rm D}$.
These  constraints, together with the small selected ratios $\sigma_{\rm CC}/\sigma_{\rm DD}=
\sigma_{\rm AA}/\sigma_{\rm BB} = 1.1$ avoid crystallization in the limit of small size disparity,
and moreover guarantee that only very small dynamic differences 
are induced between particles within a same set (\{A,B\} or \{C,D\}) \cite{mixturepaper}.
Hence, in the following we will only report dynamic quantities for A- and D-particles.

Temperature $T$, distance, wavevector $q$, 
and time $t$, will be given, respectively, in units of $\epsilon/k_B$, $\sigma_{\rm DD}$,
$\sigma_{\rm DD}^{-1}$, and $\sigma_{\rm DD}(m/\epsilon)^{1/2}$.
The packing fraction is defined as:
\begin{equation}
\phi = \frac{\pi}{6L^3}[N_{\rm A}\sigma_{\rm AA}^3 + N_{\rm B}\sigma_{\rm BB}^3 
+ N_{\rm C}\sigma_{\rm CC}^3 + N_{\rm D}\sigma_{\rm DD}^3]
\label{eq:phi}
\end{equation}
with $L$ the side of the simulation box. We fix a value $\phi = 0.53$ for
all the investigated disparities.
The system is prepared by placing the particles
randomly in the simulation box, with a constraint that avoids core overlapping.
Periodic boundary conditions are implemented.
Equations of motion are integrated by using the velocity Verlet scheme \cite{frenkel},
with a time step ranging from $2 \times 10^{-4}$ to $5 \times 10^{-3}$,
for respectively the highest and the lowest investigated temperature.
A link-cell method \cite{frenkel} is used for saving computational time
in the determination of particles within the cutoff distance of a given one. 

At each state point, the system is thermalized at the requested temperature by periodic velocity rescaling.
After reaching equilibrium, energy and pressure show no drift.
Likewise, mean squared displacements and dynamic correlators show no aging, i.e., no time shift 
when being evaluated for progressively longer time origins. 
Once the system is equilibrated, a microcanonical run is performed for production
of configurations, from which static structure factors, mean squared displacements,
and dynamic correlators are computed. For each state point, the latter quantities are averaged over
typically 20-40 independent samples.

\begin{center}
\bf{III. SIMULATION RESULTS}
\end{center}

\begin{center}
\bf{a. Static structure factors}
\end{center}

\begin{figure}
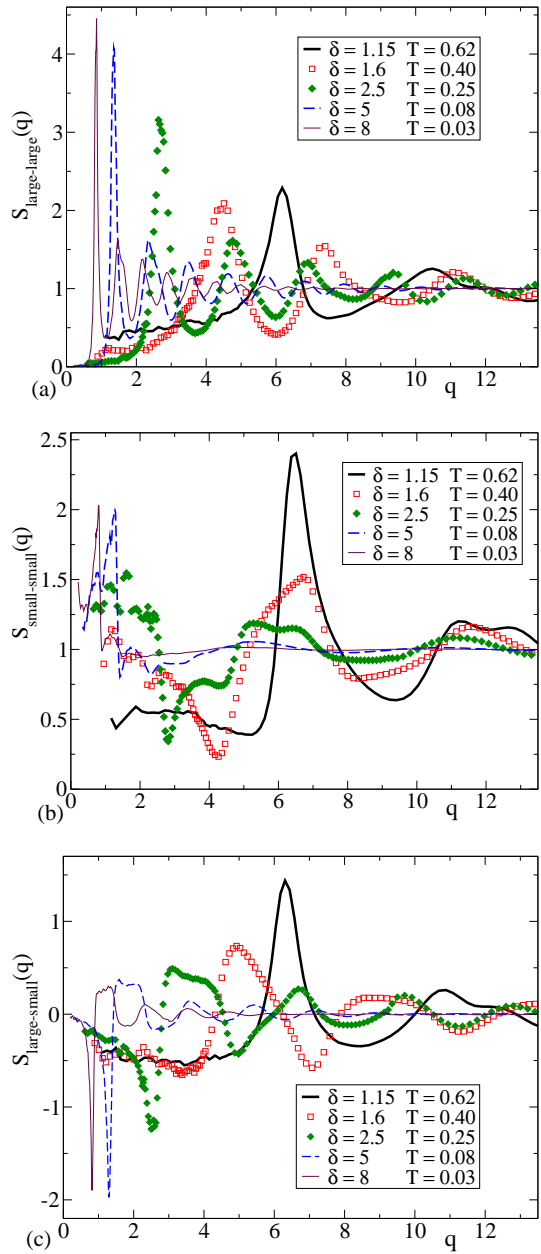

\includegraphics[width=0.86\linewidth]{505641JCP1a.eps}
\newline
\newline
\includegraphics[width=0.86\linewidth]{505641JCP1b.eps}
\newline
\newline
\includegraphics[width=0.86\linewidth]{505641JCP1c.eps}
\newline
\caption{(color online) Partial static structure factors, as computed from simulation data,
for different size disparities. Wavevectors are given in units of $\sigma_{\rm DD}^{-1}$.} 
\label{figsq}
\end{figure}

We compute normalized partial static structure factors, 
$S_{\rm \alpha\beta}(q) = \langle \rho_{\alpha}(q,0)\rho_{\beta}(-q,0)\rangle / \sqrt{N_{\alpha}N_{\beta}}$,
with $\rho_{\alpha}(q,t) = \Sigma_{j}\exp[i{\bf q}\cdot {\bf r}_{\alpha,j}(t)]$, 
the sum extending over all the particles of the species $\alpha$.
Fig. \ref{figsq} shows, for different values of $\delta$, results for
large-large, small-small, and large-small pairs. We include all the A- and B-particles 
in the set ``large'' and all the C- and D-particles in the set ``small''.
We present, for each value of $\delta$, data at the lowest investigated temperature.

$S_{\rm large-large}(q)$ exhibits a main peak at $q \approx 2\pi(\delta\sigma_{\rm DD})^{-1}$. 
This peak narrows and grows up by increasing $\delta$, reflecting a major ordering
of the large particles \cite{notecrys,dijkstra,dinsmore}. 
Likewise, the low-$q$ structure observed for small disparity
dissapears. This structure originates from the presence of inhomogeneities or ``voids'' in
the matrix of large particles, which are filled by the small particles. 
By increasing the size disparity,
the total packing fraction is largely dominated by the contribution of the large particles,
and the inhomogeneities progressively vanish.

Increasing the size disparity induces opposite effects
in $S_{\rm small-small}(q)$. The main peak observed at $q \approx 2\pi\sigma_{\rm DD}^{-1}$
for $\delta = 1.15$ is strongly damped by increasing $\delta$, reflecting a progressive
loss of the short-ranged correlations for the small particles. For large disparities,
$S_{\rm small-small}(q)$ shows a nearly structureless profile for $q > 2\sigma_{\rm DD}^{-1}$,
close to the flat behavior expected for a gas. However, a peak arises at low $q$-values,
suggesting the formation of some large-scale connected structure for the small particles.
The presence of this structure should lead to an anti-correlation between large and small particles
at the same wavevector, which is indeed manifested by a sharp negative peak in $S_{\rm large-small}(q)$.
These structural features are somewhat resembling of the observation
of channels for preferential transport of alkaline ions in silica \cite{jund,horbach,meyer}. 
As we will show below, dynamic quantities reported here also present analogies with
observations for the latter system.

\begin{center}
\bf{b. Mean-squared displacements \\ and dynamic correlators}
\end{center}

\begin{figure}
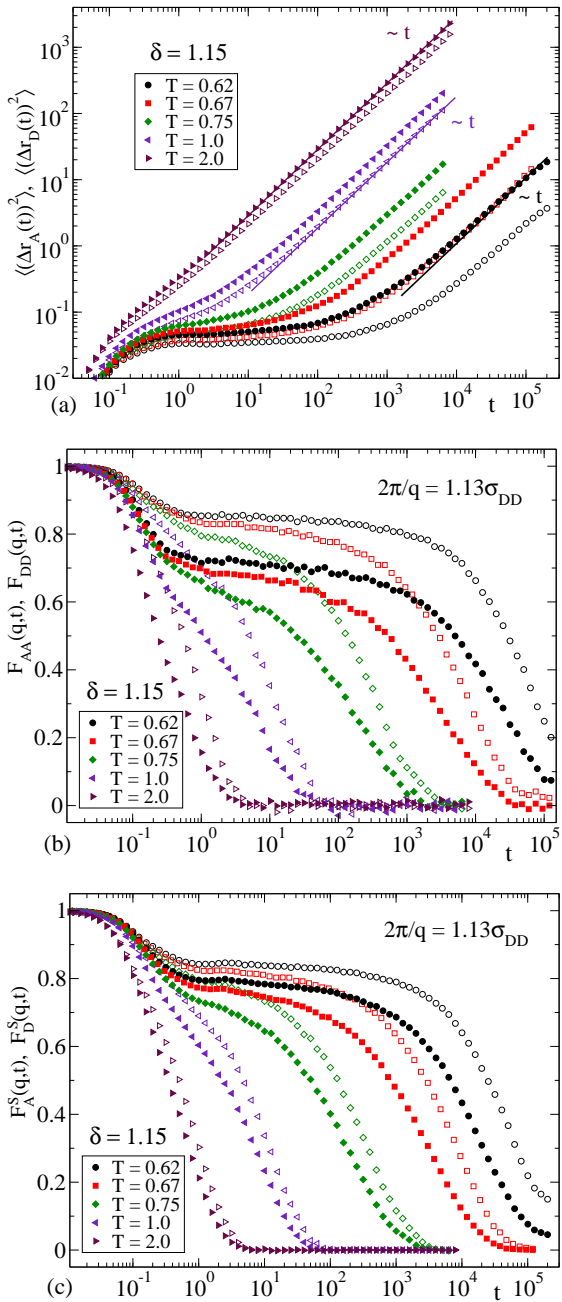

\includegraphics[width=0.9\linewidth]{505641JCP2a.eps}
\newline
\newline
\includegraphics[width=0.9\linewidth]{505641JCP2b.eps}
\newline
\newline
\includegraphics[width=0.9\linewidth]{505641JCP2c.eps}
\newline
\caption{(color online) Symbols: simulation results for size disparity $\delta = 1.15$
at different temperatures. Two identical symbols (empty and filled for respectively
A- and D-particles) correspond to a same temperature. 
Panel (a): mean-squared displacements; panel (b): density-density correlators;
panel (c): self-correlators. The two latters are plotted at a fixed wavevector 
$q = 5.56\sigma_{\rm DD}^{-1}$. Straight lines in panel (a) correspond to linear behavior.}
\label{figdel1.15}
\end{figure}

Fig. \ref{figdel1.15} displays, for size disparity $\delta = 1.15$, the temperature
dependence of mean-squared displacements (MSD, $\langle [\Delta r_{\alpha}(t)]^2 \rangle$),
density-density, and self-correlators for $\alpha =$ A- and D-particles. 
Density-density correlators for $\alpha$-$\alpha$ pairs are calculated as 
$F_{\alpha\alpha}(q,t)=\langle \rho_{\alpha}(q,t)\rho_{\alpha}(-q,0)\rangle/
\langle \rho_{\alpha}(q,0)\rho_{\alpha}(-q,0)\rangle$. Self-correlators
are computed as  $F^{\rm s}_{\alpha}(q,t) = 
\Sigma_{j}\exp\{i{\bf q}\cdot [{\bf r}_{\alpha,j}(t)-{\bf r}_{\alpha,j}(0)]\}$.
The introduction of a small size disparity induces a small separation
between the time scales of large and small particles. Only at the lowest investigated temperature,
$T = 0.62$, this separation is about a factor 5 in diffusivity (see Fig. \ref{figdel1.15}a).
As usually observed in the proximity of liquids-glass transitions \cite{koband,kobrev},
a bending occurs in the MSD after the initial ballistic ($\propto t^2$) regime. 
A plateau arises at low temperatures. This effect corresponds to the onset of the caging regime. 
At long times, the diffusive regime ($ \propto t$) 
is reached for values $\langle (\Delta r_{\alpha})^2 \rangle \lesssim \sigma_{\rm \alpha\alpha}^2$, 
i.e, when the particles have moved, on average, a distance of the order of their size.

\begin{figure}
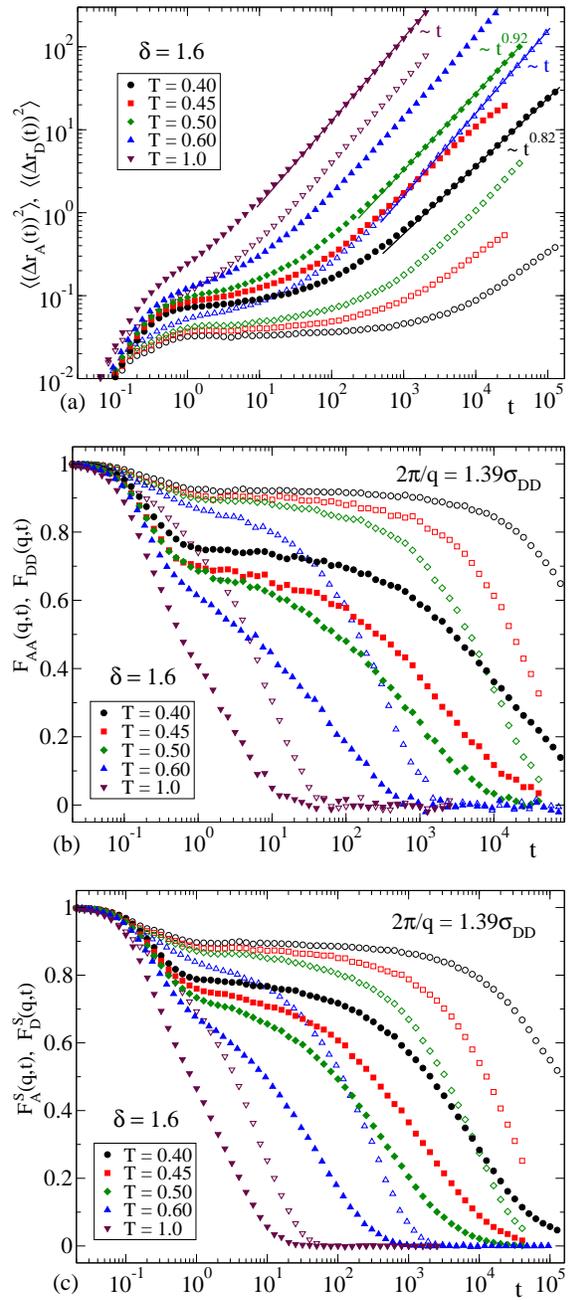

\includegraphics[width=0.9\linewidth]{505641JCP3a.eps}
\newline
\newline
\includegraphics[width=0.9\linewidth]{505641JCP3b.eps}
\newline
\newline
\includegraphics[width=0.9\linewidth]{505641JCP3c.eps}
\newline
\caption{(color online) As Fig. \ref{figdel1.15} for $\delta = 1.6$.
The wavevector for panels (b) and (c) is $q = 4.52\sigma_{\rm DD}^{-1}$.
Straight lines in panel (a) correspond to linear or sublinear power-law
behavior (exponents are given).}
\label{figdel1.6}
\end{figure}

\begin{figure}
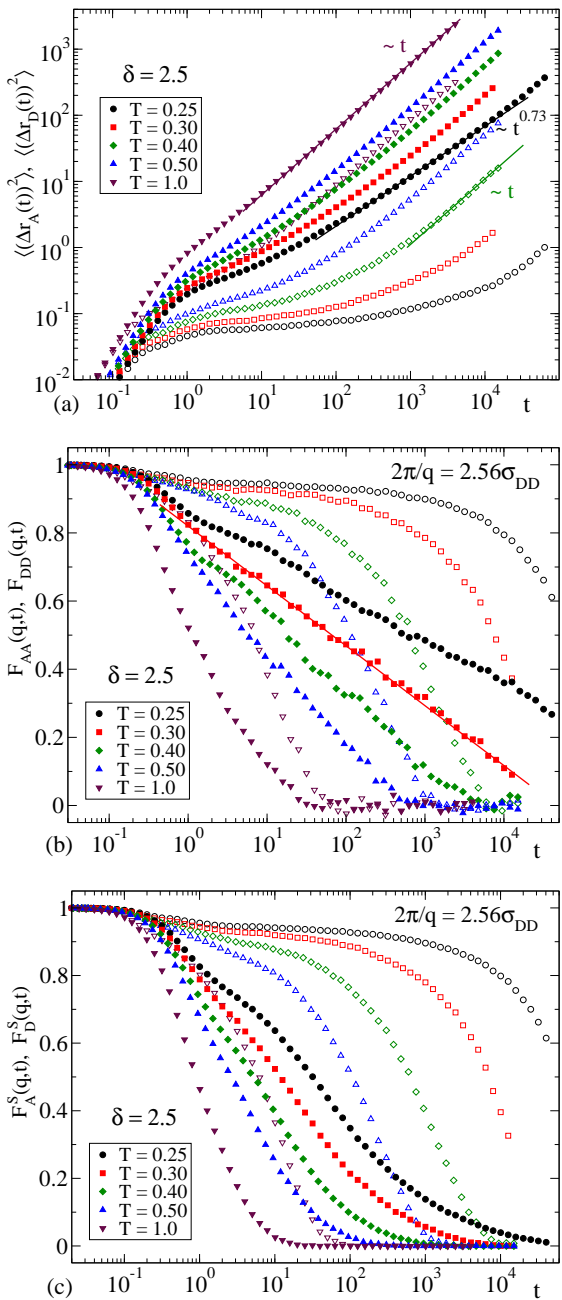

\includegraphics[width=0.9\linewidth]{505641JCP4a.eps}
\newline
\newline
\includegraphics[width=0.9\linewidth]{505641JCP4b.eps}
\newline
\newline
\includegraphics[width=0.9\linewidth]{505641JCP4c.eps}
\newline
\caption{(color online) As Fig. \ref{figdel1.6} for $\delta = 2.5$.
The wavevector for panels (b) and (c) is $q = 2.45\sigma_{\rm DD}^{-1}$.
The straight line in panel (b) indicates logarithmic-like behavior.}
\label{figdel2.5}
\end{figure}

As usual \cite{koband,kobrev,mctrev1,mctrev2}, another plateau is observed
for density-density and self-correlators (Figs. \ref{figdel1.15}b and \ref{figdel1.15}c)
in the time interval corresponding to the caging regime.
This interval is known as the $\beta$-regime within the framework of the MCT.
The correlators start to decay from the plateau at times corresponding to the onset
of the diffusive regime in the MSD. This second decay is known as the $\alpha$-regime,
and it is often described by an empirical Kohlrausch-Williams-Watt (KWW) function, given by
a stretched exponential, $A_q\exp[-(t/\tau_q)^{\beta_q}]$, with $A_q < 1$ the plateau height
and  $0 < \beta_{q} < 1$. The parameters $A_q$, $\tau_q$ and $\beta_{q}$ are $q$-dependent.

Fig. \ref{figdel1.6} displays results for disparity $\delta = 1.6$, which induces a clear
separation between the time scales for A- and D-particles. 
Interestingly, the MSD for D-particles shows, at sufficiently
low temperatures, an apparent sublinear power-law behavior, $\propto t^\alpha$, over time intervals
of almost two decades after the caging regime. Hence, differently for the standard behavior,
the small particles need to move distances of a few times their size to reach the diffusive regime. 
The exponent $\alpha < 1$ seems to decrease by decreasing temperature. 
All these features suggest that increasing the size disparity
induces a progressive separation in the time scales of the large and small particles,
and a qualitative change in the relaxation scenario for the small particles.

\begin{figure}
\includegraphics[width=0.9\linewidth]{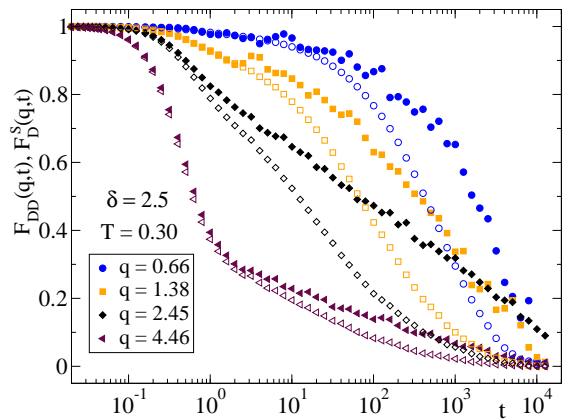}
\newline
\caption{(color online)  Dynamic correlators
of D-particles, at different wavevectors, for size disparity $\delta = 2.5$ and $T = 0.30$.
Empty and filled symbols correspond, respectively, to self- and density-density correlators.}
\label{figdel2.5T0.30}
\end{figure}

\begin{figure}
\includegraphics[width=0.9\linewidth]{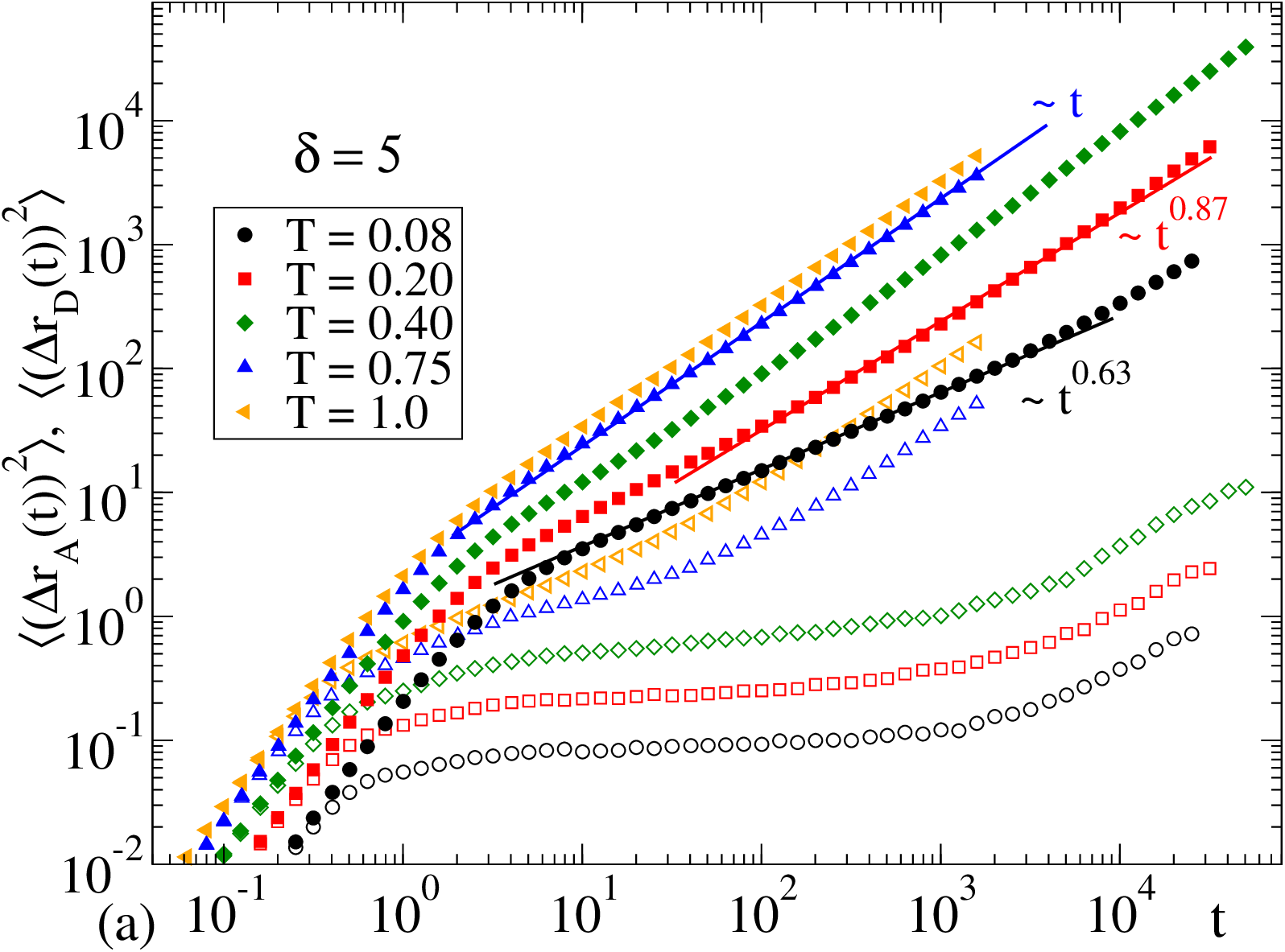}
\newline
\newline
\includegraphics[width=0.9\linewidth]{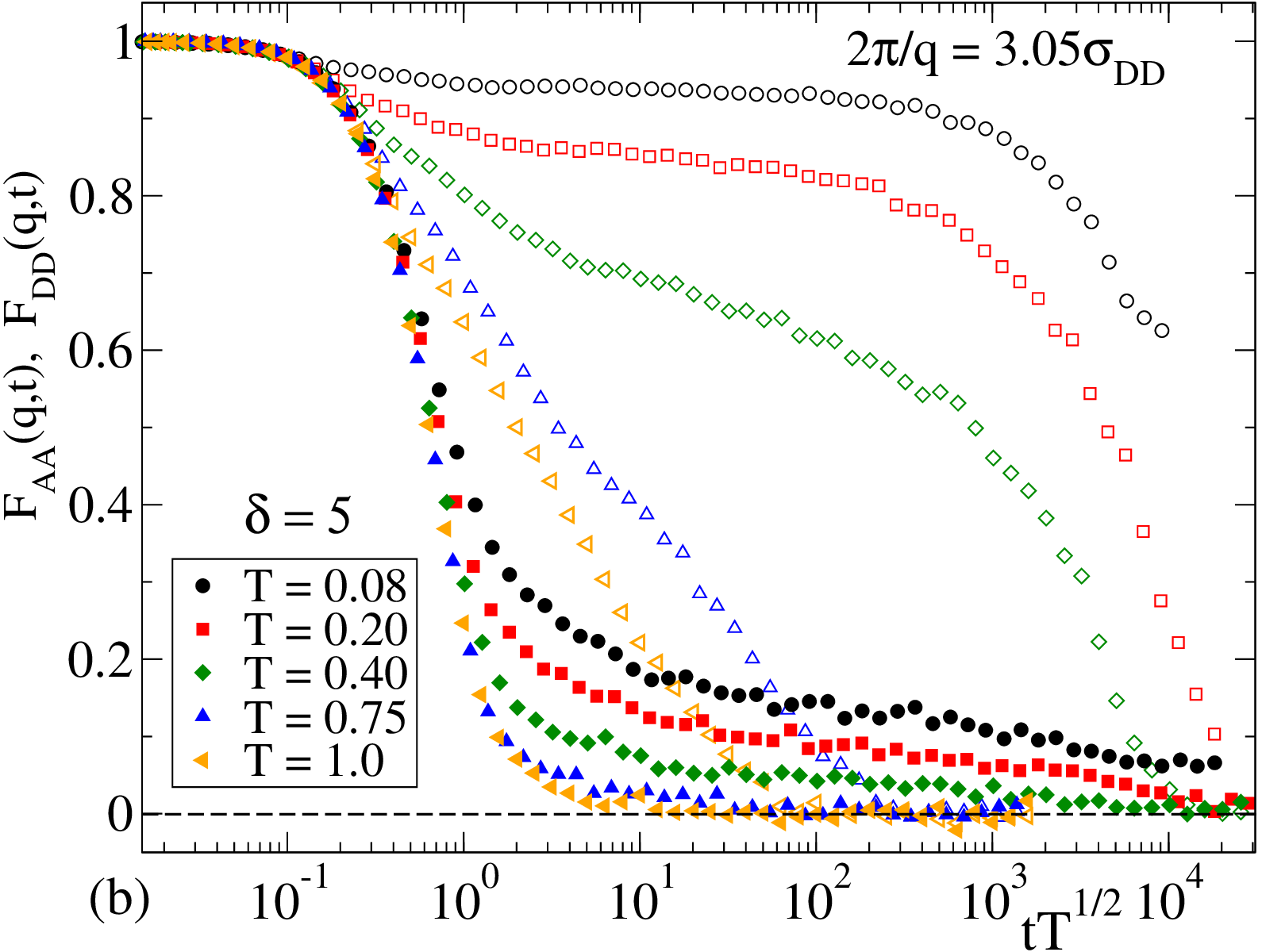}
\newline
\newline
\includegraphics[width=0.9\linewidth]{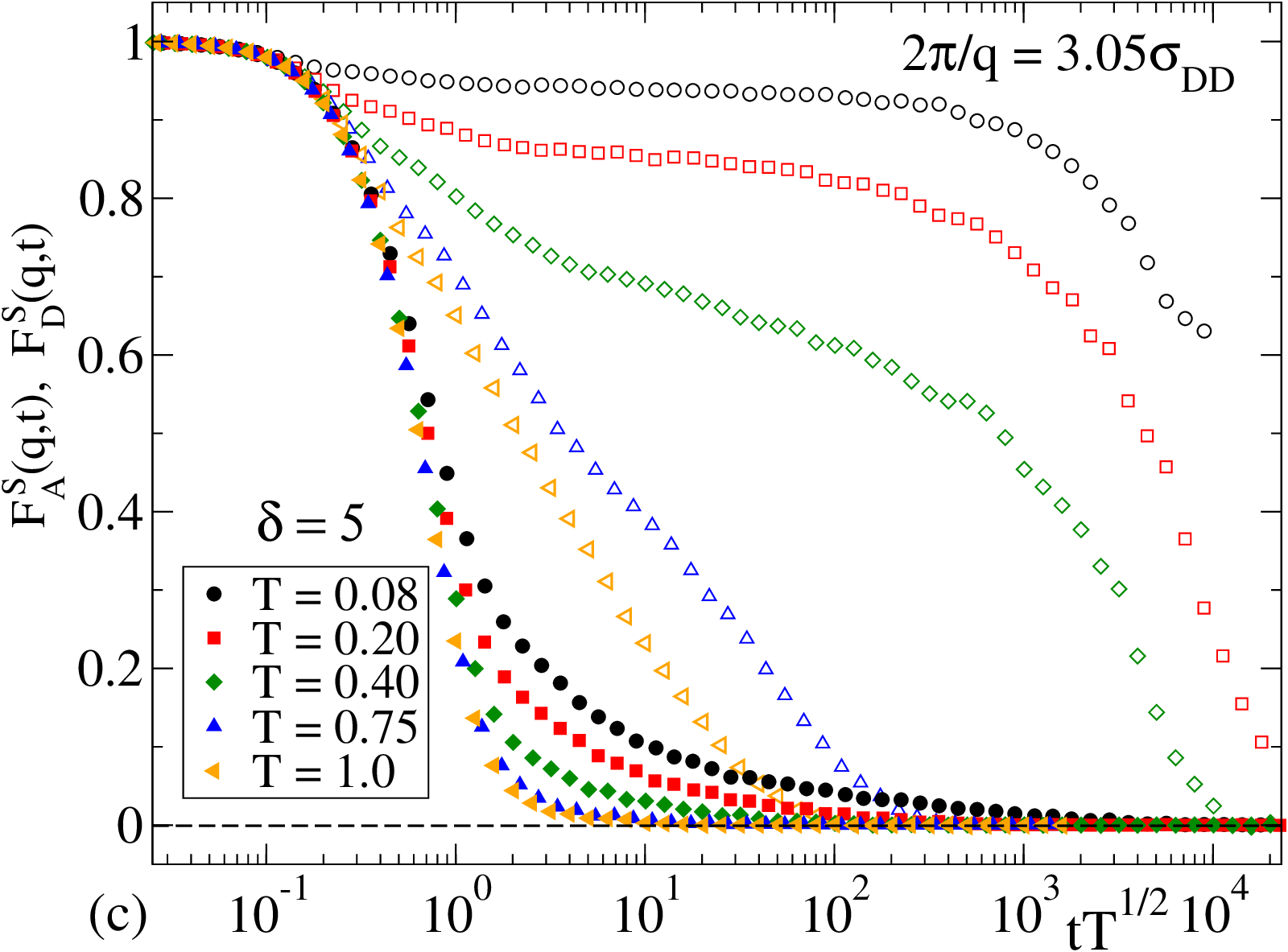}
\newline
\caption{(color online) As Fig. \ref{figdel1.6} for $\delta = 5$.
The wavevector for panels (b) and (c) is $q = 2.06\sigma_{\rm DD}^{-1}$.
Time in panels (b) and (c) is rescaled by $T^{1/2}$. The horizontal dashed lines
indicate the zero value.}
\label{figdel5.0}
\end{figure}

These trends are confirmed by results for $\delta = 2.5$ (see Fig. \ref{figdel2.5}).
D-particles reach the diffusive regime only for displacements of about ten times their size. 
Density-density correlators for the D-particles exhibit extreme stretching.
No defined plateau is present. At the lowest investigated temperatures a logarithmic decay 
is observed over a time window of about four decades. Self-correlators for the D-particles show
a qualitatively different behavior. They exhibit a decay much faster than density-density
correlators at the same temperature. This feature is displayed in Fig. \ref{figdel2.5T0.30} 
by comparing, at a fixed temperature $T = 0.30$ and several wavevectors, results
for $F_{\rm DD}(q,t)$ and $F_{\rm D}^{\rm s}(q,t)$. Both correlators only converge in
the limit of high-$q$. The time scale separation at moderate and low wavevectors
reaches values of even two decades. It is interesting to note that,
after a fast decay down to $F_{\rm D}^{\rm s}(q,t) \sim 0.3$, self-decorrelation 
is considerably slowed down, as evidenced by the long tail observed at long times
(Fig. \ref{figdel2.5}c). This tail is observed in the same time scale where the MSD
exhibits power-law subdiffusive behavior, suggesting a common microscopic origin for both features.

\begin{figure}
\includegraphics[width=0.9\linewidth]{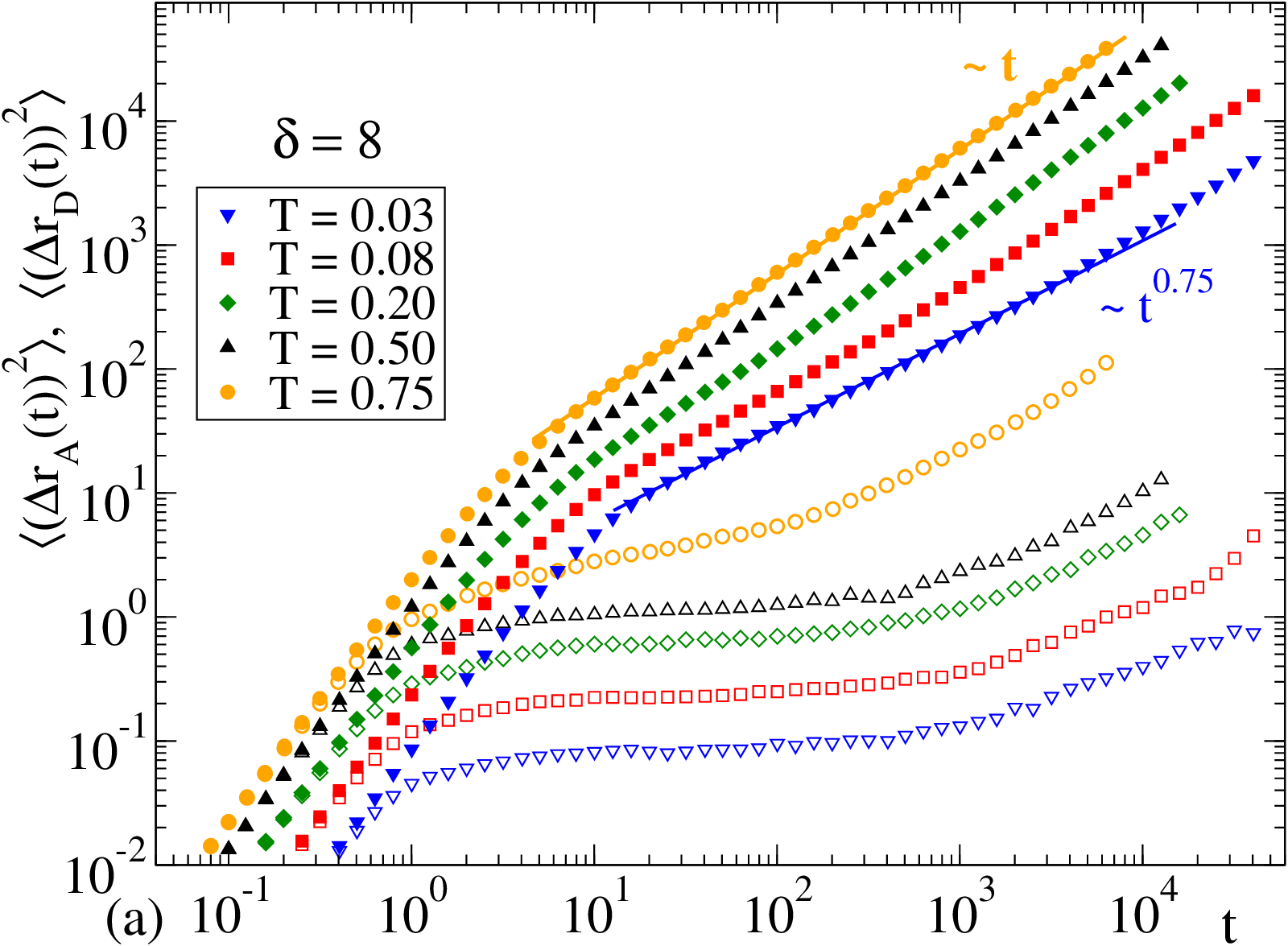}
\newline
\newline
\includegraphics[width=0.9\linewidth]{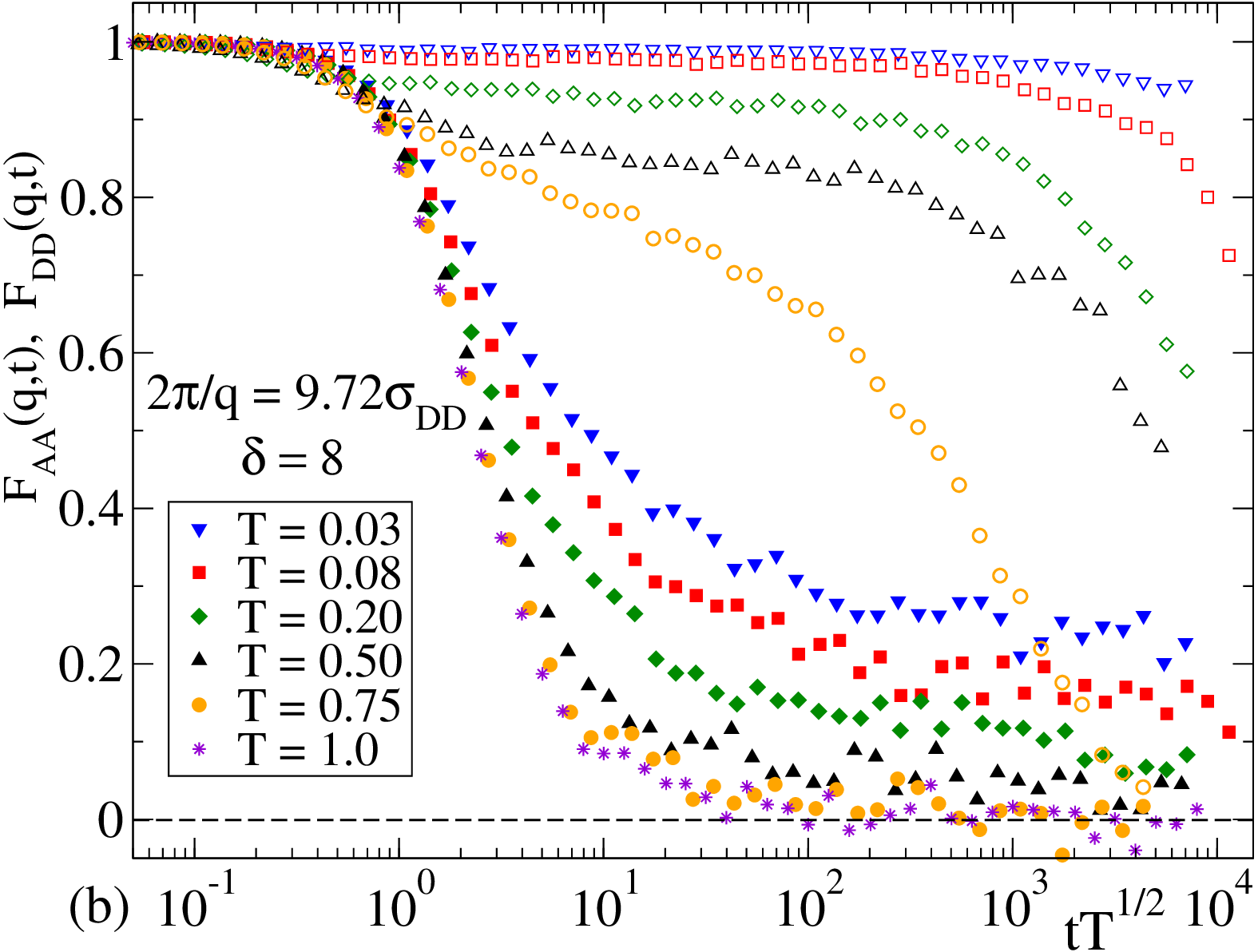}
\newline
\newline
\includegraphics[width=0.9\linewidth]{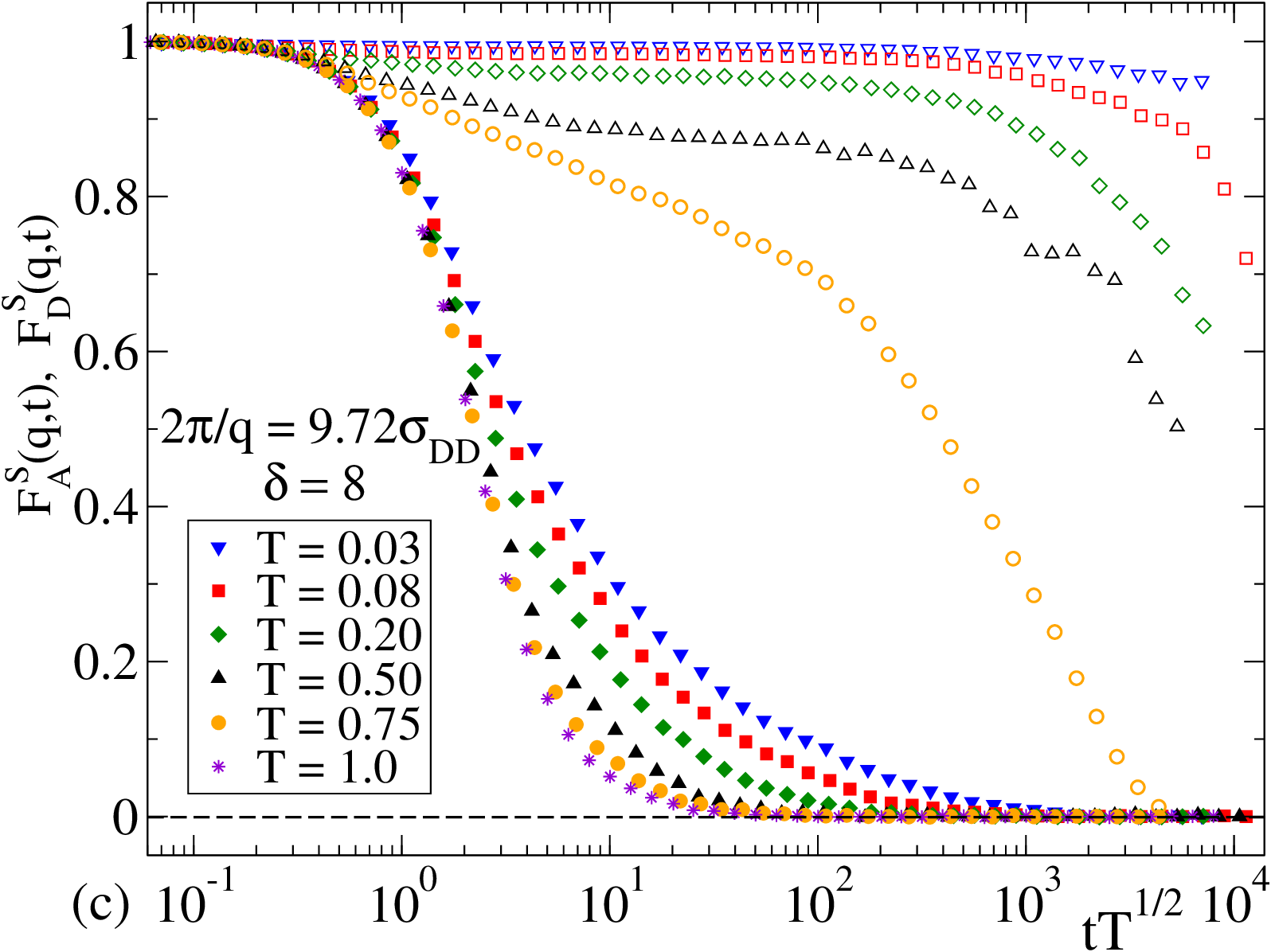}
\newline
\caption{(color online) As Fig. \ref{figdel5.0} for $\delta = 8$.
The wavevector for panels (b) and (c) is $q = 0.65\sigma_{\rm DD}^{-1}$.}
\label{figdel8.0}
\end{figure}

A new relaxation scenario arises by increasing the size disparity to much larger values.
Fig. \ref{figdel5.0} displays simulation results for $\delta = 5$. In order to provide
a clearer visualization of the decay for density-density and self-correlators, 
data for the latters are plotted as a function of $tT^{1/2}$, i.e., rescaling the time by the
thermal velocity. As expected, the large particles again exhibit a much slower dynamics.
Density-density correlators for the small particles do not show a slow decay at any temperature,
even for wavevectors corresponding to wavelengths of several particle diameters. 
Decorrelation occurs in an essentially exponential way down to very small values of $F_{\rm DD}(q,t)$, 
where a nearly flat background arises below some temperature and extends to very long times.
This feature is progressively enhanced by increasing $\delta$, as shown in Fig. \ref{figdel8.0}
for $\delta = 8$. 
The amplitude of the background increases as the system is cooled down. 
Qualitatively similar results are observed for other wavevectors. 

Self-correlators for the small particles display a striking result. As for density-density correlators, 
they exhibit a fast decay followed by a long tail. 
However, they decay to zero at all the investigated temperatures (Figs. \ref{figdel5.0}c and \ref{figdel8.0}c),
even at those where $F_{\rm DD}(q,t)$ shows a finite long-time limit. Results for the MSD in 
Figs. \ref{figdel5.0}a and \ref{figdel8.0}a evidence that the small particles reach the long-time diffusive regime
at such temperatures. All these features demostrate that the self-motion of small particles is ergodic at all 
the investigated temperatures. A small particle can reach regions arbitrarely far from its initial position.
However, according to results in Figs. \ref{figdel5.0}b and \ref{figdel8.0}b, coherent dynamics are non ergodic
below some given temperature. 

This unusual feature and its precursor effects shown above for $\delta = 2.5$
can be tentatively assigned to the existence of preferential paths for the motion 
of the small particles, forming a slowly relaxing large-scale structure. As a consequence, small
particles would be able to diffuse along such a structure, but the associated density fluctuations would
be long-lived, leading to very slow coherent dynamics as compared to self-motions.
At very large size disparities the matrix of large particles becomes glassy and the paths
for diffusion of the small particles would not be able to relax, leading to non-ergodic coherent dynamics
while self-motions remaining ergodic. Though the characterization of this eventual structure of 
preferential paths for the small particles is beyond the scope of this article, the presence of low-$q$ peaks 
in the partial static structure factors  (Figs. \ref{figsq}b and \ref{figsq}c) supports this hypothesis.
It is worth mentioning that a time scale-separation for incoherent and coherent dynamics
similar to that reported here for $\delta = 2.5$ has been observed for the diffusion of alkaline
ions in silica matrixes \cite{horbach}. In this system a channel-like structure for preferential
diffusion of the alkaline ions has been explicitely characterized \cite{jund,horbach,meyer}.
The separation of incoherent and coherent dynamics for the formers has been assigned to the
exsitence of this structure.


Results reported in this section evidence that increasing the size disparity yields unusual
relaxation scenarios for the dynamics of the small particles. In the next section we discuss 
these results within the framework of the MCT and propose a global physical picture for
the different scenarios observed at different disparities.

\begin{center}
\bf{IV. DISCUSSION}
\end{center}

\begin{center}
\bf{a. MCT framework}
\end{center}

Next we summarize the basic predictions of MCT and test them in the present system. 
In its ideal version, which neglects activated hopping events, MCT predicts
a sharp transition from an ergodic liquid to a non-ergodic arrested state (glass) at a given value
of the relevant control parameter (in the following the temperature, $T$) 
\cite{mctrev1,mctrev2,das}. When crossing the transition point $T=T_{\rm c}$ (critical temperature)
from the ergodic to the arrested state, the long-time limit of the correlator for wavevector $q$, $F(q,t)$,
jumps from zero to a non-zero value, denoted as the {\it critical} non-ergodicity parameter, $f^{\rm c}_q$. 
Moving beyond the transition point into the non-ergodic state yields a progressive increase 
of the non-ergodicity parameter, $f_q > f^{\rm c}_q$. 
In the standard case the jump in $F(q,t)$ is discontinuous, i.e., the critical non-ergodicity 
parameter $f^{\rm c}_q$ takes a finite value.
MCT transitions with $f^{\rm c}_q > 0$ are denoted as {\it type-B} transitions.
For ergodic states close to the transition point, the initial part of the $\alpha$-process
 --- i.e., the von Schweidler regime --- is given by a power law expansion \cite{mctrev1,mctrev2,das}:
\begin{equation}
F(q,t) \approx f^{\rm c}_q -h_q (t/\tau_{\rm c})^{b} + h_q^{(2)}(t/\tau_{\rm c})^{2b},
\label{eqvonsch}
\end{equation}
with $0 < b \le 1$. The prefactors $h_q$ and $h_q^{(2)}$ only depend on
the wavevector $q$ and are different for each correlator. 
The characteristic time $\tau_{\rm c}$ only depends on the separation parameter $|T - T_{\rm c}|$
and is divergent at the transition point. 
As mentioned above, the decay from the plateau to zero is often described by an empirical
KWW function with a $q$-dependent stretching exponent $\beta_q$. An interesting prediction of MCT
\cite{fuchs} is that $\beta_q = b$ in the limit of large-$q$. This result has been widely tested
\cite{koband,horbach,bennemann,vanzon,starr,mossa,puertaspre,gallofq} 
and provides a consistency test for data analysis.

\begin{figure}
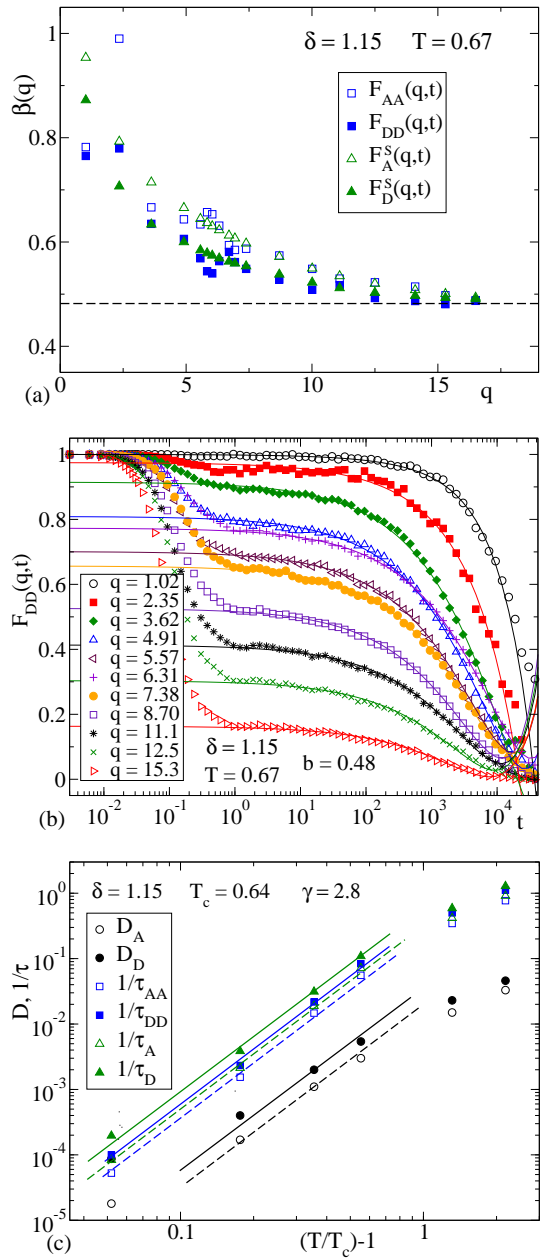

\includegraphics[width=0.86\linewidth]{505641JCP8a.eps}
\newline
\newline
\includegraphics[width=0.86\linewidth]{505641JCP8b.eps}
\newline
\newline
\includegraphics[width=0.86\linewidth]{505641JCP8c.eps}
\newline
\caption{(color online) MCT analysis for $\delta = 1.15$. Symbols are simulation results.
Panel (a): For $T = 0.67$, $q$-dependence of the stretching exponents obtained from fitting the decay 
from the plateau to a KWW function. The corresponding correlators are given in the legend.
The dashed line indicates the limit $\beta (q \rightarrow \infty ) \approx 0.48$.
Panel (b): density-density correlators for D-D pairs, at $T = 0.67$, for different wavevectors.
Curves are fits to Eq. (\ref{eqvonsch}) with $b = 0.48$.
Panel (c): diffusivities and inverse relaxation times for $q = 6.3\sigma_{\rm DD}^{-1}$ 
(see text for the definition).
Empty and filled symbols correspond, respectively, to A- and D-particles. Dashed and solid straight lines
are fits to Eq. (\ref{eqpower}) for, respectively, A- and D-particles. The values $T_{\rm c} = 0.64$
and $\gamma = 2.8$ are forced. }
\label{figmctdel1.15}
\end{figure}

\begin{figure}
\includegraphics[width=0.9\linewidth]{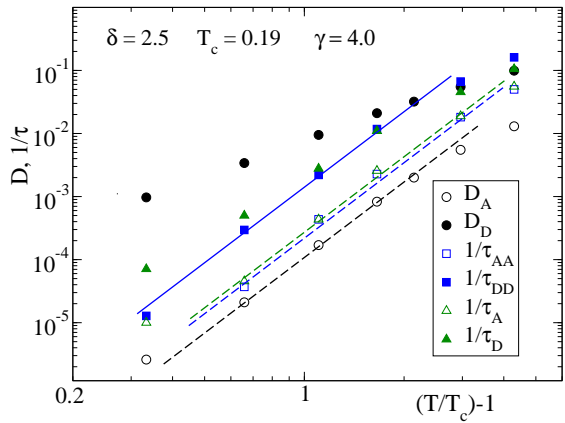}
\newline
\caption{(color online) As Fig. \ref{figmctdel1.15}c for $\delta = 2.5$. The values $T_{\rm c} = 0.19$
and $\gamma = 4.0$ are forced. The wavevector for the $\alpha$-relaxation
times is $q = 2.45\sigma_{\rm DD}^{-1}$.}
\label{figmctdel2.5}
\end{figure}

Another prediction of MCT for state points close to the transition point
is the power law dependence of the diffusivity and the $\alpha$-relaxation time:

\begin{equation}
D , 1/\tau_{\alpha} \propto (T-T_{\rm c})^{\gamma}
\label{eqpower}
\end{equation}
The $\alpha$-relaxation time, $\tau_{\alpha}$, is a time scale probing the $\alpha$-process.
In practice it can be defined as the time $\tau_{x}$ where $F(q,t)$ decays to some small value $x$, 
provided it is well below the plateau.
The exponent $\gamma$ in Eq. (\ref{eqpower}) is given by \cite{mctrev1,mctrev2,das}:
\begin{equation}
\gamma = \frac{1}{2a} + \frac{1}{2b},
\label{eqgamma}
\end{equation}
with $0 < a < 0.4$. Hence $\gamma \ge 1.75$. The critical exponents $a$, $b$, and $\gamma$
are univoquely related with the so-called exponent parameter $1/2 \le \lambda \le 1$ through:
\begin{equation}
\lambda = \frac{\Gamma^{2}(1+b)}{\Gamma(1+2b)} = \frac{\Gamma^{2}(1-a)}{\Gamma(1-2a)},
\label{eqlambda}
\end{equation}
with $\Gamma$ the Gamma function \cite{mctrev1,mctrev2,das}. The exponent parameter $\lambda$
is univoquely determined by the static correlations (i.e., by the total and partial static structure factors)
at the transition point $T = T_{\rm c}$. 

When numerical solutions of the MCT equations are not available
the non-ergodicity parameters, prefactors and exponents 
in Eqs. (\ref{eqvonsch},\ref{eqpower},\ref{eqgamma},\ref{eqlambda}) 
are obtained as fit parameters from simulation or experimental data. 
Consistency of the data analysis requires that the so-obtained values for the exponents fulfill
Eqs. (\ref{eqgamma}) and (\ref{eqlambda}).
It must be stressed that values of MCT critical exponents obtained as fit parameters must be taken
with care, since the range of validity (in time and temperature windows) 
of the {\it asymptotic} Eqs. (\ref{eqvonsch},\ref{eqpower}) is not known {\it a priori}.
The case of the power law (\ref{eqpower}) is specially problematic, not
only for the upper limit in temperature, but also for the lower one. Deviations of the power-law
behavior are often observed below some ill-defined temperature very close
to $T_{\rm c}$ \cite{ashwin,gallo,flenner},
due to the presence of activated hopping events which are not accounted by ideal MCT. 
Hence, by selecting different temperature ranges (both the upper and lower limits being ill-defined)
for fits to Eq. (\ref{eqpower}) one can find different {\it effective} exponents $\gamma$, which
can be rather different from the actual one. It often occurs that the effective exponents obtained for the
diffusivities are very different from those obtained for the relaxation times \cite{koband,gallofq,horbachsil}.  
See, e.g., Refs. \cite{flenner,chong} for a discussion on this point.

Fig. \ref{figmctdel1.15} shows a test of MCT predictions for size disparity $\delta =1.15$.
Within numerical uncertainties, the stretching exponents of self- and density-density correlators, 
both for large and small particles, seem to approach a common value $\beta \approx 0.48$ in the large-$q$ limit
(Fig. \ref{figmctdel1.15}a), in agreement with MCT expectations \cite{fuchs}. 
Consequently with the prediction $\beta (q \rightarrow \infty) = b$, 
we fix the von Schweidler exponent $b$ to this value and fit, for the different
correlators, the initial decay from the plateau to Eq. (\ref{eqvonsch}). Consistently, a good description
of the simulation data is achieved. Fig. \ref{figmctdel1.15}b shows the corresponding fits for $F_{\rm DD}(q,t)$.

The value $b = 0.48$ provides through Eqs. (\ref{eqgamma}) and (\ref{eqlambda}) the values
$\lambda = 0.79$, $a = 0.28$, and $\gamma = 2.8$. Now we test the validity of Eq. (\ref{eqpower}) with
this latter value of $\gamma$.
Fig. \ref{figmctdel1.15}c shows fits to Eq. (\ref{eqpower}) of diffusivities and $\alpha$-relaxation times. 
The diffusivities $D_{\alpha}$ are calculated as the long-time limit
of $\langle [\Delta r_{\alpha}(t)]^2 \rangle/6t$. We use $x = 0.2$ for the definition
of the $\alpha$-relaxation times $\tau_x$ (denoted as $\tau_{\alpha}^{\rm s}$ 
for self-correlators and $\tau_{\alpha\alpha}$ for density-density correlators). The latters are shown
for $q = 6.3\sigma_{\rm DD}^{-1}$.  We have fixed $b = 0.48$ and forced a {\it common} $T_{\rm c}$ for all
the diffusivities and relaxation times, as predicted by MCT for monodisperse spheres 
or binary mixtures with small size disparity \cite{gotzevoigtmann}. 
The best fits are achieved with $T_{\rm c} = 0.64$. As expected, the power law (\ref{eqpower})
fails above some high temperature. For the diffusivities it also fails at temperatures very close to $T_{\rm c}$,
where activated hopping events are expected to raise the simulation values above the theoretical predictions.
It is often found that these deviations arise for diffusivities at higher temperatures than
for relaxation times \cite{flenner}. This is not surprising, since diffusivities and relaxation times are dominated
by the contribution of, respectively, fast and slow particles. Similarly, deviations due to
hopping events in binary mixtures are stronger for the small particles \cite{gallofq,gallo,flenner}, 
which are, on average, faster than the large ones. This feature is indeed observed in Fig. \ref{figmctdel1.15}c
and is enhanced by increasing $\delta$ (see Fig. \ref{figmctdel2.5} for $\delta = 2.5$),
suggesting a major relevance of hopping events for the diffusivity of the small particles at larger disparity.
As discussed above, a fully free fit to Eq. (\ref{eqpower}) would provide rather 
different exponents $\gamma$ for relaxation times and diffusivities. Only a simultaneous analysis 
of dynamic correlators, as illustrated in Figs. \ref{figmctdel1.15}a and \ref{figmctdel1.15}b can provide 
a robust determination of the critical exponents.

An analogous analysis for disparity $\delta = 1.6$ provides for the critical exponents the values $a = 0.25$, 
$b = 0.40$, $\gamma = 3.2$, and $\lambda = 0.85$. We obtain a critical temperature $T_{\rm c} = 0.42$,
much lower than that for $\delta = 1.15$. Hence, increasing the size disparity to $\delta = 1.6$
at constant packing fraction shifts the glass transition to lower temperatures, 
as predicted by MCT \cite{gotzevoigtmann} for similar values of $\delta$ and observed
in experiments \cite{williams} and simulations \cite{foffiprl,foffipre}.

\begin{figure}
\includegraphics[width=0.9\linewidth]{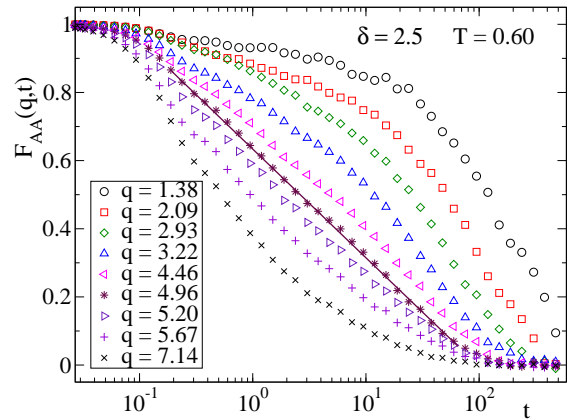}
\newline
\caption{(color online) Density-density correlator for A-A pairs at disparity $\delta = 2.5$
and temperature $T = 0.60$. Different symbols correspond to different wavevectors.
The straight line indicates logarithmic decay.}
\label{figfaaT0.60del2.5}
\end{figure}

This trend is also fulfilled for $\delta = 2.5$, for which the critical temperature is $T_{\rm c} = 0.19$.
Fig. \ref{figmctdel2.5} shows simulation results for the diffusivities and the relaxation times.
Due to the small amplitude of the tail exhibited by the self-correlators for the D-particles
(see Fig. \ref{figdel2.5}c), we have defined the corresponding relaxation time $\tau_x$ by using $x = 0.03$.
The value $x = 0.2$ is used for the other relaxation times displayed in Fig. \ref{figmctdel2.5}.
We obtain the values $a = 0.21$, $b = 0.31$, $\gamma = 4.0$, and $\lambda = 0.90$ 
for the critical exponents \cite{noteexp}.
Another predicted trend \cite{gotzevoigtmann} that is fulfilled by increasing $\delta$ is the observed
increase of the exponent parameter $\lambda$. 

In our previous article (see Fig. 9 in Ref. \cite{mixturepaper}) we have analyzed
the decay of $F_{\rm DD}(q,t)$ at $\delta = 2.5$ in terms of a logarithmic expansion,
\begin{equation}
F(q,t) \approx f^{\rm c}_q -H_q \ln (t/\tau_{\rm c}) + H_q^{(2)}\ln^2 (t/\tau_{\rm c}),
\label{eqlog}
\end{equation}
instead of the von Schweidler law (\ref{eqvonsch}). Within the framework of MCT, Eq. (\ref{eqlog}) is associated
to a nearby higher-order transition \cite{schematic,dawson,sperl}, which is characterized
by a value of the exponent parameter $\lambda = 1$, or to a standard transition
with a value of $\lambda$ very close to 1. Hence, the value $\lambda = 0.90$ here obtained 
for disparity $\delta = 2.5$ justifies the description used in Ref. \cite{mixturepaper} in terms
of a logarithmic expansion.

It must be stressed that, in principle, this choice is not in contradiction with the description, for
this same value of $\delta = 2.5$, of the
correlator for the A-particles in terms of a von Schweidler law, 
as done in Ref. \cite{mixturepaper} (see Fig. 7b there).
According to MCT the presence of a nearby higher-order transition is manifested in a given
correlator $F(q,t)$ by a purely logarithmic decay at some given temperature. Moreover, by varying $q$ at that
temperature, the shape of the decay exhibits a convex-to-concave crossover 
\cite{sperl,dawson,prlsimA4,zaccarelli1}. We have observed these features for the small particles
at low temperatures (see Fig. 9a in Ref. \cite{mixturepaper}), but they are absent for the large particles
at the same temperatures (Fig. 7b in Ref. \cite{mixturepaper}).
However, they are clearly observed at higher temperatures, as shown here in Fig. \ref{figfaaT0.60del2.5}
for $T = 0.60$. Logarithmic relaxation covers more than two time decades at $q \approx 5\sigma_{\rm DD}^{-1}$.
The fact that these anomalous features (logarithmic relaxation and convex-to-concave crossover)
are observed for the large and the small particles at different temperatures is not, in principle, 
a failure of MCT. The optimal distance to the transition point
for the observation of anomalous relaxation is determined by the coefficient $H_q^{(2)}$
in Eq. (\ref{eqlog}). This coefficient is decomposed \cite{sperl} as
$H_q^{(2)}=A(\{{\bf x}^n\}) + B(\{{\bf x}^n\})K_q$,
where the $q$-independent terms $A(\{{\bf x}^n\})$ and $B(\{{\bf x}^n\})$
depend on the state point $\{{\bf x}^n\} = \{T, \delta, x_{\rm small}, \phi,...\}$ in the control parameter space,
and $K_q$, which is determined by static correlations, only depends on $q$ and is different for each correlator. 
As a result of this decomposition, for a given correlator there are paths
in the control parameter space where $H_q^{(2)} = 0$ and along which, according to Eq. (\ref{eqlog}), a purely
logarithmic decay will be observed. The fact that in the present case the partial static structure factors for
the large and the small particles are very different (Fig. \ref{figsq}) 
might yield very different $K_q$-coefficients for the respective correlators, and as a consequence, 
very different paths for purely logarithmic relaxation of the large and the small particles.
This feature might explain why anomalous relaxation for both types of partices is observed at very
different temperatures. Unless one follows the optimal path in the control parameter space --- which in principle
involves a simultaneous variation of {\it several} control parameters \cite{schematic,sperl}--- 
anomalous relaxation vanishes by decreasing temperature, and a standard two-step decay is recovered
(see Ref. \cite{zaccarelli1} for an illustrative example).
Hence, correlators for the large particles at low temperature are well described by a von Schweidler law.
The fact that anomalous relaxation is present for the small particles at very low temperatures
suggests that decreasing temperature and fixing the other control parameters
does not deviate the system far from the corresponding optimal path for the small particles.

\begin{figure}
\includegraphics[width=0.86\linewidth]{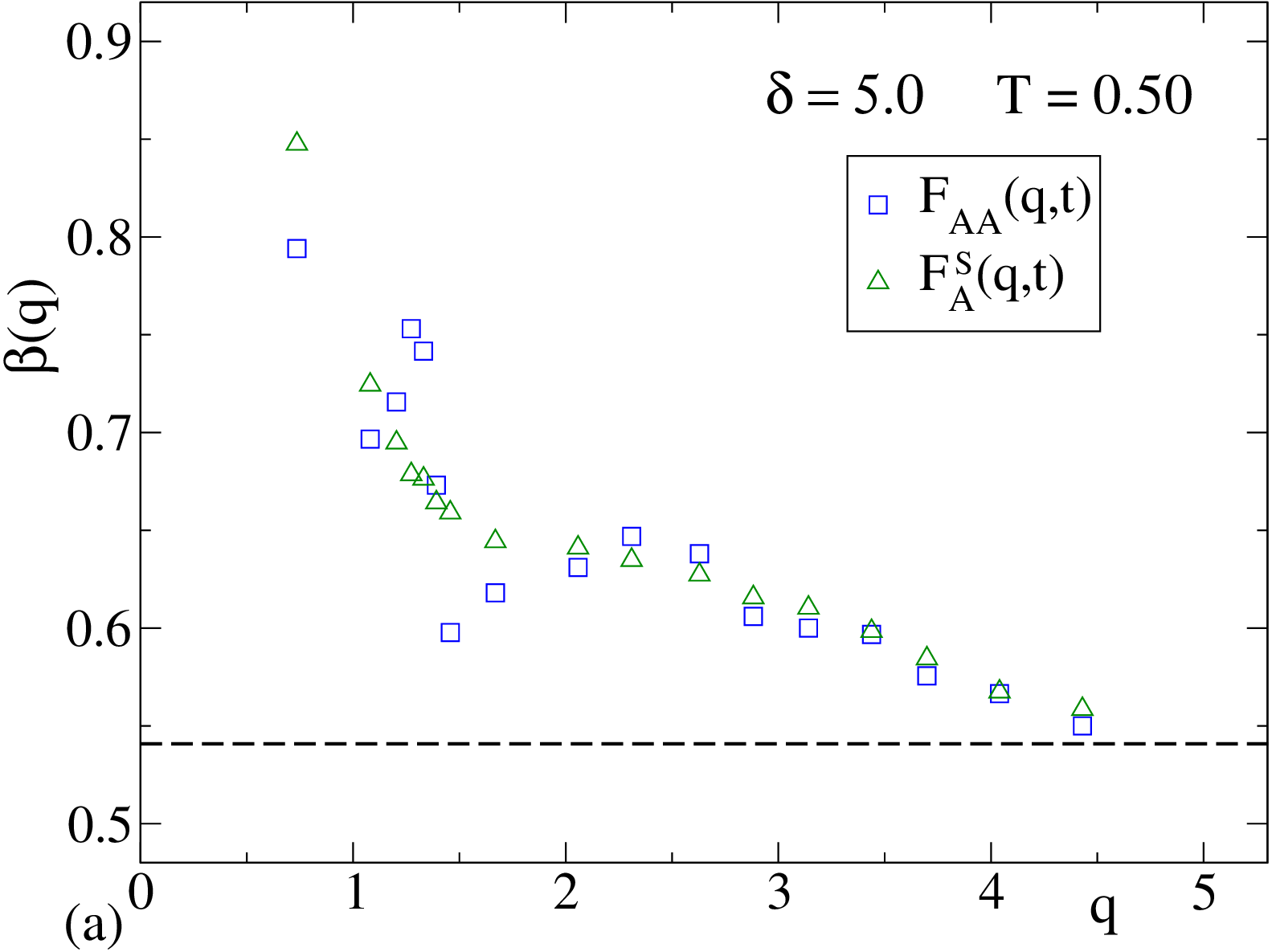}
\newline
\newline
\includegraphics[width=0.86\linewidth]{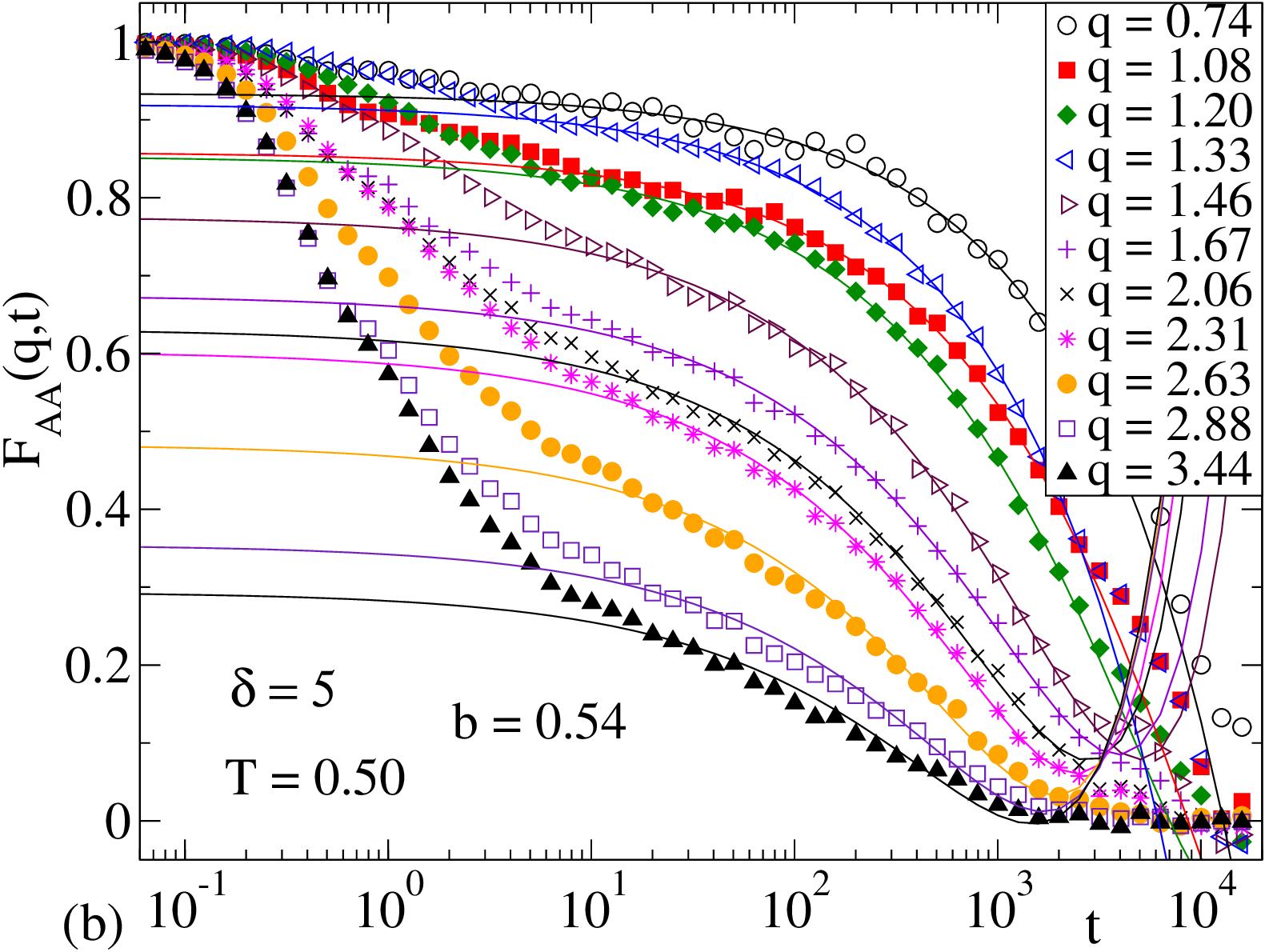}
\newline
\newline
\includegraphics[width=0.86\linewidth]{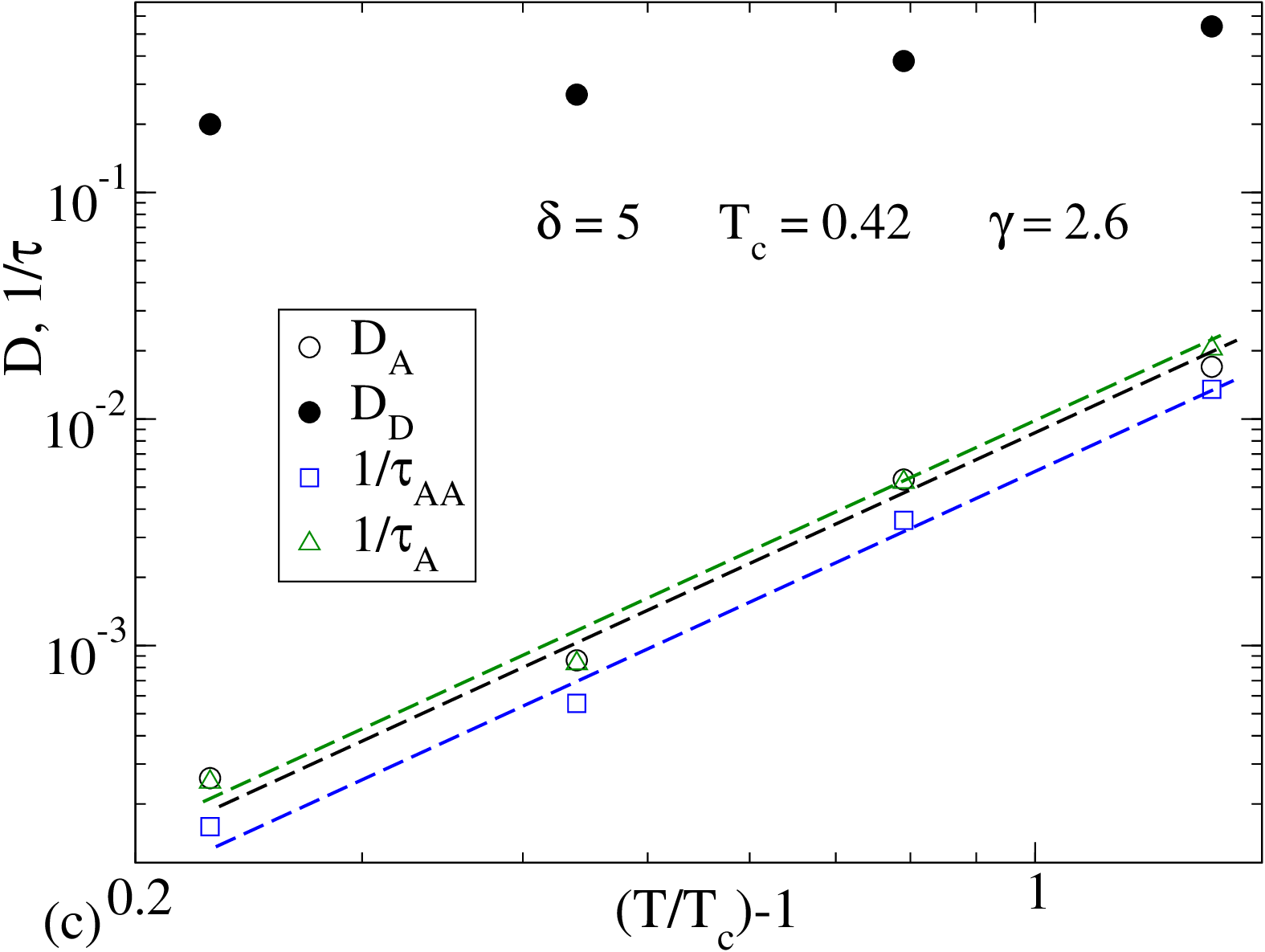}
\newline
\caption{(color online) MCT analysis for $\delta = 5$. Symbols are simulation results.
Panel (a): For $T = 0.50$, $q$-dependence of the stretching exponents obtained from fitting the decay 
from the plateau to a KWW function. The corresponding correlators are given in the legend.
The dashed line indicates the limit $\beta (q \rightarrow \infty ) \approx 0.54$.
Panel (b): density-density correlators for A-A pairs, at $T = 0.50$, for different wavevectors.
Curves are fits to Eq. (\ref{eqvonsch}) with $b = 0.54$.
Panel (c): diffusivities and inverse relaxation times for $q = 1.33\sigma_{\rm DD}^{-1}$.
Empty and filled symbols correspond, respectively, to A- and D-particles. Dashed lines
are fits to Eq. (\ref{eqpower}). The values $T_{\rm c} = 0.42$ and $\gamma = 2.6$ are forced.}
\label{figmctdel5.0}
\end{figure}

The analysis for the {\it large} particles at $\delta = 5$ (see Fig. \ref{figmctdel5.0})
provides a critical temperature $T_{\rm c} = 0.42$. Hence, the trend observed for smaller disparities is reversed
for sufficiently large values of $\delta$, and the glass transition exhibits a reentrant behavior 
by using $\delta$ as a control parameter. 
Increasing disparity at constant packing fraction initally decreases $T_{\rm c}$. However, the latter reaches some
minimal value and beyond some $\delta$ it starts to increase. 
For $\delta = 5$ we obtain the critical exponents $a = 0.30$, $b = 0.54$, $\gamma = 2.6$, and $\lambda = 0.76$.
Hence, the reentrant behavior observed for $T_{\rm c}$ also involves a non-monotonic
behavior for the exponent parameter $\lambda$. Increasing $\delta$ initially raises $\lambda$
to large values (0.90 for $\delta = 2.5$), suggesting an eventual nearby higher-order transition
($\lambda = 1$). Beyond some disparity, for which it reaches a maximum,  
the exponent parameter starts to decrease. These trends are confirmed for the largest investigated
disparity $\delta = 8$, for which we obtain $T_{\rm c} = 0.52$ and the critical exponents 
$a = 0.31$, $b = 0.57$, $\gamma = 2.5$, and $\lambda = 0.74$. 
We note on passing that data for correlators of the large particles
in Figs. \ref{figdel5.0} and \ref{figdel8.0} show a decay from the plateau at temperatures
far below $T_{\rm c}$. This feature is in principle related to the mentioned hopping events
not included in the ideal version of MCT, which restore ergodicity below $T_{\rm c}$.
The fact that, despite the final decay, the plateau height exhibits a clear increase below the value
estimated for $T_{\rm c}$ is consistent with expectations of ideal MCT.

\begin{figure}
\includegraphics[width=0.9\linewidth]{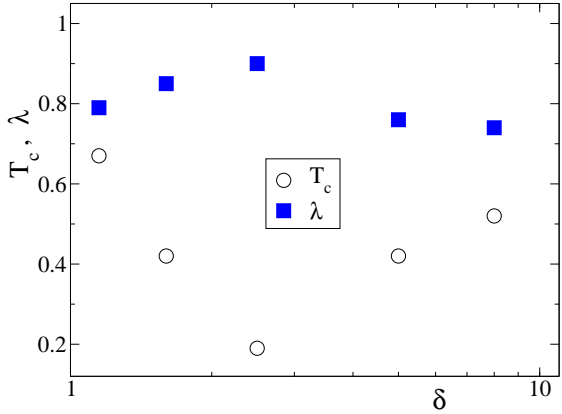}
\newline
\caption{(color online) Circles: Critical temperatures $T_{\rm c}$ for the different investigated
disparities. Squares are the corresponding values of the exponent parameter $\lambda$. 
Note the logarithmic scale in $\delta$. Error bars are given by the symbol size.}
\label{figTclambda}
\end{figure}

Fig. \ref{figTclambda} displays the values of $T_{\rm c}$ and $\lambda$ obtained at the
different investigated disparities. Only small variations of these values
provide reasonable descriptions of simulation results in a consistent way (i.e., by fitting
simultaneously different data sets as exposed, e.g., in Fig. \ref{figmctdel1.15}).
The so-estimated error bars for $T_{\rm c}$ and $\lambda$
are given by the symbol size in Fig. \ref{figTclambda}.
The values of $\lambda$ obtained for $ \delta \le 1.6$
are somewhat higher than those predicted by MCT \cite{gotzevoigtmann} in the same range for
a binary mixture of hard spheres at the same composition $x_{\rm small} = 0.6$. 
From Fig. 13 in Ref. \cite{gotzevoigtmann} (note the transformations of $x_{\rm small}$ 
and $\delta$ to the quantities represented there), we estimate $\lambda \approx 0.73$ 
for $\delta = 1.15$ and $\lambda \approx 0.78$ for $\delta = 1.6$. 
Unfortunately, values of $\lambda$ for larger disparities have not been reported.

It must be stressed that determining wether there is actually
a higher-order point ($\lambda = 1$) at some disparity around $\delta = 2.5$, or wether $\lambda$
instead reaches a maximum $0.90 \le \lambda_{\rm max} < 1$, is a difficult task. 
Since, from Eq. (\ref{eqlambda}), $\lambda \rightarrow 1$
implies that $b \rightarrow 0$, a determination of the specific value of $b$ through the limit
$b = \beta (q \rightarrow \infty)$ is not reliable for extremely small values of $b$.
Only the solution of the MCT equations can unambiguously determine the actual value of $\lambda_{\rm max}$.
Still, the existence or absence of a higher-order point is not a question so important as the anomalous relaxation
features associated to it (logarithmic decays and convex-to-concave crossover in dynamic correlators, as well as 
sublinear power-law behavior for the MSD \cite{sperl}), which are also expected for values $\lambda \rightarrow 1$
as those found here.

For the case of the {\it small} particles at disparities $\delta = 5$ and $\delta = 8$ we have not performed
an analysis in terms of MCT similar to those above exposed. Indeed, data for the small particles
in Figs. \ref{figdel5.0} and \ref{figdel8.0} exhibit a very different scenario of
that observed for $\delta \le 2.5$. As mentioned in Section III, density-density
correlators do not show a slow decay, but a fast one followed by a nearly flat background below some temperature,
its amplitude growing up by decreasing temperature. We identify the background amplitude as an operational 
non-ergodicity parameter that increases from some critical value $f_q^{\rm c}$.
As shown in Figs. \ref{figdel5.0}b and \ref{figdel8.0}b it is difficult 
to establish wether $f_q^{\rm c}$ defined in this way is zero, but it is clear 
that it takes, as much, a extremely small value. We can also define an operational $T_{\rm c}$ 
for the {\it small} particles as the temperature where $f_q^{\rm c}$ starts to increase.
Though a precise determination of this temperature is difficult, simulation
data suggest that $0.4 \lesssim T_{\rm c} \lesssim 0.75$ and $0.5 \lesssim T_{\rm c} \lesssim 0.75$
for, respectively, $\delta = 5$ and $\delta = 8$, i.e., of the order of the critical
temperature for the {\it large } particles (see above), and compatible 
with a common $T_{\rm c}$ for all the particles.

Data for $\delta = 5$  show a final decay of the background to zero (see Fig. \ref{figdel5.0}b). 
This decay is, again, presumably related with the presence of hopping events. Note that despite of this decay,
the increase of the background amplitude for $T \lesssim 0.4$ is evident, as expected for the
crossing of a MCT critical temperature.

The behavior reported here for density-density correlators of the small particles at $\delta = 5$
and $\delta = 8$ resembles features characteristic of MCT transitions of the so-called type-A, which are defined
by a {\it zero} value of the critical non-ergodicity parameter, in contrast to the finite value defining
the standard type-B transitions. A recent realization of this scenario has been reported 
for a system of dumbbell molecules \cite{chong,dumbbellsprl,dumbbellsjcp}.
While for moderate molecular elongations a standard relaxation scenario is observed, 
for small elongations angular correlators of odd order exhibit features analogous to those
of Figs. \ref{figdel5.0}b and \ref{figdel8.0}b \cite{dumbbellsjcp}. Theoretical calculations 
for this latter system \cite{chong} relate such features to the existence 
of a MCT transition of the type-A.

Data presented here are consistent with early MCT theoretical calculations by Bosse
and co-workers \cite{bosse,kaneko} and later by Harbola and Das \cite{harbola}
for a binary mixture of hard spheres with the packing fraction as the control parameter. 
Though in these works no information is given about relaxation features, calculations are reported
for non-ergodicity parameters of self- ($f^{\rm s}_q$) and density-density ($f_q$) correlators. 
These calculations predict, for sufficiently large disparities, a glassy phase for density-density correlations
($f_q > 0$) where self-correlations for the small particles remain ergodic ($f^{\rm s}_q = 0$) \cite{bosse}
and the diffusivity for the latters is finite \cite{kaneko}, in agreement with simulation results presented
in this article and with early experiments by Imhof and Dhont \cite{imhof} in binary mixtures 
of colloidal silica particles with size disparity $\delta = 9.3$. While density-density correlators of the small
particles display a critical temperature $T_{\rm c} > 0.4$, 
Figs. \ref{figdel5.0}c and \ref{figdel8.0}c do not show, even for $T \ll T_{\rm c}$,
any signature of a crossing of a critical temperature for self-correlators.
Likewise, small particles exhibit large values of the diffusivities 
($D_{\rm D} > 0.02$ for $\delta = 8$) and very weak caging effects in the MSD 
(Figs. \ref{figdel5.0}a and \ref{figdel8.0}a) at temperatures $T \ll T_{\rm c}$.

As mentioned in Section III, transport of alkaline ions in silica display a separation between the time
scales for coherent and incoherent dynamics similar to that presented here for $\delta = 2.5$.
In a recent work, MCT calculations by Voigtmann and Horbach have reproduced this feature \cite{voigtmannhorbach},
which has also been interpreted as a precursor effect of an eventual type-A transition \cite{procmainz}.
Given the similarities (already noted in \cite{procmainz}) of this system with binary mixtures
of sufficiently large size disparity, the former might be a candidate to display, 
by properly tuning the concentration of alkaline ions, many of the unusual relaxation features reported here.
Another possible candidate is a mixture of star polymers with arms of very different length.
In a coarse-grained picture, this system can be seen as binary mixture of 
{\it ultrasoft} spheres \cite{starpol,likos} with large size disparity. 
By properly tuning the mixture composition and the disparity it is possible to obtain a glassy matrix
formed by the large stars where the small ones remain fluid \cite{starpol}.

\begin{figure}
\includegraphics[width=0.86\linewidth]{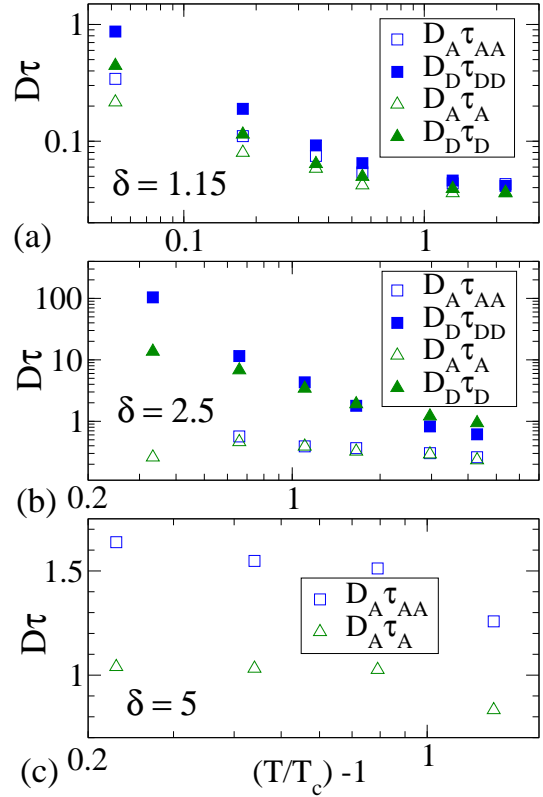}
\newline
\caption{(color online) For different size disparities, product of diffusivities $D$
by $\alpha$-relaxation times $\tau$. The values of the latter quantities, 
as well as the critical temperatures $T_{\rm c}$, 
are the same ones as in Figs. \ref{figmctdel1.15}c, \ref{figmctdel2.5}, and \ref{figmctdel5.0}c.
All the axes are in logarithmic scale, except for the vertical one in panel (c), 
which is in linear scale. }
\vspace{4 mm}
\label{figDtau}
\end{figure}

To finish this subsection, we make a brief comment on MCT predictions
for $D\tau$, the  product of the diffusivities by the $\alpha$-relaxation time. 
Fig. \ref{figDtau} shows, for different size disparities, results for $D\tau$
as computed from numerical data of Figs. \ref{figmctdel1.15}c, \ref{figmctdel2.5}, 
and \ref{figmctdel5.0}c. This product approaches a constant value at high temperatures and, for some
of the data sets in Fig. \ref{figDtau}, increases by orders of magnitude
by decreasing temperature. This feature has been widely observed in very different
systems \cite{mossa,puertaspre,horbachsil,chong,foffipre,flennerdtau,voigtmannpuertas}
and is usually attributed to the presence of dynamic heterogeneities,
which enhance the diffusivity (dominated by the contribution of the fast particles)
as compared to the $\alpha$-relaxation time (dominated by the contribution of the slow particles). 
It must be emphasized that deviations of $D\tau$ from constant behavior are
not {\it a priori} a signature of deviations from MCT predictions. Since the power law 
$D \propto (T-T_{\rm c})^{\gamma}$ is just a {\it leading-order} result, MCT {\it does not}
predict a temperature independent behavior for the product $D\tau$, but just a weak variation
between the high-$T$ limit and a bounded value at $T \rightarrow T_{\rm c}$ 
(see e.g., Refs.\cite{chong,voigtmannpuertas}). This prediction is compatible with data for
the large particles at $\delta = 2.5$ (Fig. \ref{figDtau}b) and $\delta = 5$ (Fig. \ref{figDtau}c).
Indeed only small deviations from MCT, in the investigated $T$-range, are observed
by looking separately at $D_{\rm A}$, $\tau_{\rm AA}$, and $\tau_{\rm A}$ 
(Figs. \ref{figmctdel2.5} and \ref{figmctdel5.0}c). Deviations for the diffusivity are larger
at $\delta = 1.15$ (Fig. \ref{figmctdel1.15}c), leading to an increase of about an order
of magnitude in $D_{\rm A}\tau_{\rm AA}$ and $D_{\rm A}\tau_{\rm A}$ within 
the investigated $T$-range (Fig. \ref{figDtau}a).
On the contrary, $D\tau$ for the small particles strongly deviates from the MCT prediction
in all cases, as expected from the strong deviations already observed 
by looking separately at $D_{\rm D}$.

\begin{center}
\bf{b. A global picture}
\end{center}

The different relaxation scenarios here reported for the small particles by varying the
size disparity provide a connection with MCT theoretical results by Krakoviack
in Refs. \cite{krakoviackprl,krakoviackjpcm} for a mixture of {\it fixed and mobile} hard spheres,
a model introduced to reproduce qualitative dynamic features of fluids confined in 
matrixes with interconnected voids. In that work, the dynamic phase diagram of the
mobile particles displays two lines of type-A and type-B transitions
in the plane $\phi_{\rm fix}$-$\phi_{\rm mob}$, where $\phi_{\rm fix}$
and $\phi_{\rm mob}$ are the packing fractions of, respectively, the fixed and mobile particles.
The A- and B-lines merge at a higher-order point (namely an A$_3$-point \cite{noteorder}).
The B-line extends from the A$_3$-point to the limit $\phi_{\rm fix} = 0$, where the fluid
of hard spheres is recovered. The A-line extends from the A$_3$-point to the Lorentz gas
limit (a single particle diffusing in a matrix of fixed obstacles) at $\phi_{\rm mob} = 0$.

For high concentrations of the mobile particles, 
the matrix of fixed particles does not yield signifficant confinement effects. The only
caging mechanism is normal hard-sphere repulsion at short length and time scales,
and dynamic correlators exhibit a standard two-step decay \cite{krakoviackjpcm}. The transition
point is of the type-B and hence the jump of the long-time limit of the density-density correlator
is finite \cite{krakoviackjpcm}, providing a non-zero value of the critical non-ergodicity parameter. 

For high dilution of the mobile particles, hard-sphere repulsion at short length and time scales
does not yield temporary caging. As a consequence, density-density correlators display a fast decay
to values close to zero \cite{krakoviackjpcm}. At longer time scales the mobile particles 
probe the structure of the confining matrix of fixed particles, which leads 
to a ``mesoscopic'' caging, characterized by a length scale
larger than that characteristic of bulk-like hard-sphere repulsion. As a consequence of this large-scale
caging mechanism (confinement), density-density correlators for the mobile particles exhibit a long tail of small
amplitude after the fast microscopic decay \cite{krakoviackjpcm}. At the transition point, of the type-A,
the long-time limit does not exhibit a finite jump but grows up continuously \cite{krakoviackjpcm},
providing a zero value for the critical non-ergodicity parameter.

As mentioned above, at moderate concentrations of fixed and mobile particles,
a higher-order A$_3$-point arises \cite{krakoviackprl}
as a consequence of the competition between the mentioned ``microscopic'' and ``mesoscopic'' caging
mechanisms. Relaxation features have not been reported in Refs. \cite{krakoviackprl,krakoviackjpcm}
for state points close to the A$_3$-point. However, as stressed in \cite{krakoviackprl},
they will {\it necessarily} display the anomalies reported here (logarithmic decay and convex-to-concave
crossover in dynamic correlators, and sublinear power-law behavior for the MSD), as a mathematical consequence 
of the value $\lambda =1$ defining the A$_3$-point \cite{notematrix}.

How do results in Refs. \cite{krakoviackprl,krakoviackjpcm} compare with relaxation features presented here?.
First it is worth emphasizing that, differently from the mobile particles in that work,
high dilution of the small particles is not the key ingredient for yielding a type-A transition 
for the latters in the system here investigated.
Indeed, data reported in our previuos article \cite{mixturepaper} 
for $\delta = 2.5$, $\phi = 0.53$, and $x_{\rm small}=0.1$
show features rather different from those characterizing type-A transitions. 
It must be noted that, for these control parameters,
the system here investigated is much denser than at state points close to the A-line for the mixture of fixed 
and mobile particles of Ref. \cite{krakoviackprl}. Despite its low density, decorrelation of the mobile
particles in the latter system is blocked {\it at large length scales} due to the absence
of percolating free volume. In the system here investigated, relaxation is possible
at higher densities due to the non-static nature of the confining matrix. 
The slow motion of the large particles creates regions of sufficient local free volume 
which facilitate decorrelation of the small particles. As a consequence of high density,
short-range bulk-like caging is a relevant arrest mechanism for small particles and leads to a slow 
decay of dynamic correlators (Fig. 10a in Ref. \cite{mixturepaper}), 
different from features characteristic of type-A transitions.
Hence, for not sufficiently large disparities as $\delta = 2.5$, type-A transitions 
cannot exist at any mixture composition.

\begin{figure}
\includegraphics[width=0.86\linewidth]{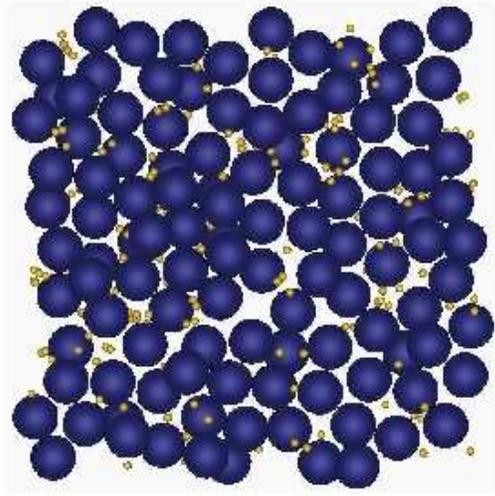}
\newline
\caption{(color online) Typical slab (of thickness $10\sigma_{\rm DD}$) 
at $T=0.03$ for size disparity $\delta = 8$.
Dark: A- and B-particles. Light: C- and D-particles.}
\vspace{4 mm}
\label{figslab}
\end{figure}

\begin{figure}
\includegraphics[width=0.9\linewidth]{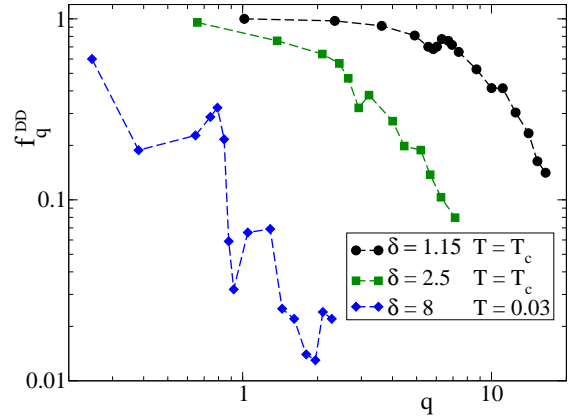}
\newline
\caption{(color online) Symbols: non-ergodicity parameters of $F_{\rm DD}(q,t)$ for different size disparities.
Results for $\delta = 1.15$ and $\delta = 2.5$ are the {\it critical} values. Results for $\delta = 8$
correspond to a state point, $T = 0.03$, below the A-line (see text).
Note the double logarithmic scale. Dashed lines are guides for the eyes.}
\label{figfqdd}
\end{figure}

A way of yielding a type-A scenario for the small particles in the present system
--- at a fixed packing fraction and mixture composition --- is by increasing the size disparity, as shown above
for $\delta = 5$ and $\delta = 8$. In this situation small particles move in a medium of low local density
and short-range bulk-like caging is supressed. This effect is illustrated in Fig. \ref{figslab}, 
which depicts a typical slab of the simulation box for $\delta = 8$. 
However, small particles are caged at large length scales where they probe the structure 
of the confining medium. This crossover from
 ``microscopic'' to ``mesoscopic'' caging for the small particles by increasing the size disparity 
is manifested by a strong narrowing of the $q$-dependence of the non-ergodicity parameter
(see Fig. \ref{figfqdd}), which is a signature of an increasing localization length \cite{bosse}. 
Below a certain temperature, where the tail of the soft-sphere potential is probed,
the effective packing of the confining matrix will become sufficiently large to freeze density-density
correlations for the small particles, leading to a dynamic picture analogous to the
type-A scenario observed for the system of Refs. \cite{krakoviackprl,krakoviackjpcm}.

Relaxation features reported here for the small particles at $\delta = 5$ and $\delta = 8$
suggest the presence of a nearby A-line, originating from large-scale caging (confinement)
induced by the slow matrix of large particles. For small size disparity, $\delta \rightarrow 1$, 
one obviously recovers the standard MCT scenario
predicted for the monoatomic fluid of hard spheres and widely observed in mixtures of
hard \cite{foffipre}, soft \cite{barrat}, or Lennard-Jones spheres \cite{koband}, where small disparity 
is just introduced as a way of avoiding crystallization. The standard scenario, originating
from short-range bulk-like caging, is characterized by a nearby B-line. 

Increasing the size disparity at a fixed mixture composition will weaken
the effects of bulk-like caging and strengthen those associated to confinement. This feature suggests 
that a crossover from a B- to an A-line will occur by increasing the value of $\delta$,
in analogy with results for the mixture of fixed and mobile spheres investigated 
in Refs. \cite{krakoviackprl,krakoviackjpcm},
where such a crossover is obtained by varying the composition.
For the latter system, the B- and A-lines merge at a higher-order point.
The resemblance of results reported here and in Ref. \cite{mixturepaper} for $\delta = 2.5$ 
with relaxation features characterizing higher-order MCT transitions
suggest an analogous merging. The observed variation of the exponent parameter $\lambda$ 
by increasing $\delta$ (Fig. \ref{figTclambda}) also supports this hypothesis, or at least, the existence
of a ``quasi'' higher-order point with $\lambda$ very close to 1. Anomalous relaxation at $\delta \sim 2.5$ 
would originate from the presence of a nearby B-line (due to the finite value of the critical 
non-ergodicity parameters), ending at a nearby  higher- (or quasi-higher) order point, 
which would produce the observed anomalous relaxation features.

\begin{center}

{\bf V. CONCLUSIONS}
\end{center}

We have carried out simulations on a mixture of large and small particles.
Slow dynamics have been investigated for a broad range of size disparities at a fixed
mixture composition, and discussed within the framework of the Mode Coupling Theory (MCT).
Results presented here extend a recent investigation
on the composition dependence at a fixed large disparity \cite{mixturepaper}.
Increasing the size disparity yields a crossover between different relaxation scenarios.
For sufficiently large disparities an anomalous relaxation
scenario arises, displaying features very different from the standard ones observed at small disparity.
Some of these features, which resemble predictions of MCT for state points close to higher-order transitions,
are sublinear power-law behavior for mean squared displacements,
and logarithmic decays and convex-to-concave crossovers for dynamic correlators.

For very large disparities, a new scenario arises, showing features associated 
to MCT transitions of the type-A, which are characterized by a zero value of the critical
non-ergodicity parameter, different from the finite value defining the B-transitions
observed at smaller disparities. In this scenario, self-correlations for the small particles
remain ergodic at temperatures far below the freezing of density-density correlations.
Small particles remain fluid in the glassy matrix formed by the large particles.

All these features provide a connection with MCT theoretical results for a mixture
of mobile and fixed particles \cite{krakoviackprl,krakoviackjpcm}, which report a dynamic phase
diagram displaying an A- and a B-line merging at a higher-order point. 
A similar crossover is suggested for the system here investigated by varying the size disparity. 
If this hypothesis is correct, the global MCT picture discussed here might not be a specific
feature of liquids confined in matrixes with interconnected voids \cite{krakoviackprl,krakoviackjpcm}, 
but a more general one. Of course, a proper answer to this question can only be provided by solving the
corresponding MCT equations. However, the highly non-trivial observed analogies
suggest to consider it as a plausible hypothesis and might motivate theoretical work in this problem
within the framework of the MCT. To our knowledge, there is only an approach different from MCT
which has been able to reproduce the basic phenomenology characterizing the ``higher order-like'' scenario. 
Namely a dynamic facilitation picture, which has reproduced reentrant behavior of diffusivities
for short-ranged attractive colloids \cite{geissler}, as well as concave-to-convex crossovers 
\cite{dynfacpaper} and logarithmic decays in dynamic correlators \cite{geissler,dynfacpaper}, 
and sublinear power-law behavior in the MSD \cite{dynfacmsd}. However, 
a similar picture is missing for the ``higher order-like'' scenario as an intermediate state
of a crossover between a ``type B-like'' and a ``type A-like'' scenario by varying the size disparity.

The mixture of soft spheres investigated in this work 
shares many dynamic features with other mixed systems of different nature in the case that the two
components show very different mobilities, as polymer blends \cite{blendpaper}, 
colloidal mixtures \cite{imhof,dinsmore}, star polymer mixtures \cite{starpol},
or ion conducting glasses \cite{horbach,voigtmannhorbach}. Such features in this
latter systems have been interpreted and/or predicted within the framework of MCT.
Simulation results in this article and in Ref. \cite{mixturepaper} by varying, respectively,
the size disparity and mixture composition, provide a global picture which might motivate 
experiments in a huge variety of systems in order to test eventual analogies and crossovers
between different relaxation scenarios as those presented here.

\begin{center}
\bf{ACKNOWLEDGEMENTS}
\end{center}
   
We acknowledge support from the projects NMP3-CT-2004-502235 (SoftComp), 
MAT2004-01017 (Spain), and 206.215-13568/2001 (GV-UPV/EHU Spain).


\begin{thebibliography}{9}
\bibitem{noteexp} The analysis presented here is more robust than that presented
by ourselves at $\delta = 2.5$ and the same composition in Ref. \cite{mixturepaper}, which 
performed less consistency tests. Values reported in this article for $T_{\rm c}$ and
for the critical exponents slightly correct the ones given there.
\bibitem{noteorder} A MCT transition of order $n$, A$_n$, is characterized by $n$ solutions
of the critical non-ergodicity parameter, only the highest value being physical.
Standard liquid-glass transitions are of order $n = 2$ (A$_2$), 
with exponent parameter $\lambda < 1$.
\bibitem{notematrix} Simulations reported in the literature for mixtures of fixed and mobile
particles \cite{gallofq,gallo} do not display the anomalous relaxation features associated to the
higher-order point. However, as discussed in Ref. \cite{mixturepaper}, the higher-order point 
is far from the region of the control parameter space explored by such simulations.
\end{thebibliography}
\end{document}